\documentclass[sigconf]{acmart}
\usepackage{amsmath}
\usepackage{graphicx}
\usepackage{multirow}
\usepackage{epstopdf}
\usepackage{wrapfig}
\usepackage{algorithm, algorithmic}
\usepackage{tikz}

\usepackage[T1]{fontenc}
\usepackage[utf8]{inputenc}

\usepackage{bm}
\usepackage{dsfont} 
\usepackage[mathscr]{euscript} 
\let\euscr\mathscr \let\mathscr\relax
\usepackage[scr]{rsfso} 
\usepackage{enumitem}
\usetikzlibrary{bayesnet}
\usepackage{enumitem}
\usepackage{caption}
\usepackage{xspace}

\def\BibTeX{{\rm B\kern-.05em{\sc i\kern-.025em b}\kern-.08emT\kern-.1667em\lower.7ex\hbox{E}\kern-.125emX}}

\setcopyright{acmcopyright}
\begin{document}

%
\title[Friend or Faux: Graph-Based Early Detection of Fake Accounts]{Friend or Faux: Graph-Based Early Detection of Fake Accounts\\ on Social Networks}
%


\author{Adam Breuer}
\email{breuer@g.harvard.edu}
\affiliation{%
  \institution{Harvard}
  \city{Cambridge}
  \state{MA}
  \country{USA}
}%


\author{Roee Eilat}
\email{reilat@fb.com}
\affiliation{%
  \institution{Facebook}
  \city{Menlo Park}
  \state{CA}
  \country{USA}
}%

\author{Udi Weinsberg}
\email{udi@fb.com}
\affiliation{%
  \institution{Facebook}
  \city{Menlo Park}
  \state{CA}
  \country{USA}
}%


%

\begin{abstract}

In this paper, we study the problem of early detection of fake user accounts on social networks based solely on their network connectivity with other users. Removing such accounts is a core task for maintaining the integrity of social networks, and early detection helps to reduce the harm that such accounts inflict. However, new fake accounts are notoriously difficult to detect via graph-based algorithms, as their small number of connections are unlikely to reflect a significant structural difference from those of new real accounts. We present the \textsc{SybilEdge} algorithm, which determines whether a new user is a fake account (`sybil') by aggregating over (I) her choices of friend request targets and (II) these targets' respective responses. \textsc{SybilEdge} performs this aggregation giving more weight to a user's choices of targets to the extent that these targets are preferred by other fakes versus real users, and also to the extent that these targets respond differently to fakes versus real users. We show that \textsc{SybilEdge} rapidly detects new fake users at scale on the Facebook network and outperforms state-of-the-art algorithms. We also show that \textsc{SybilEdge} is robust to label noise in the training data, to different prevalences of fake accounts in the network, and to several different ways fakes can select targets for their friend requests. To our knowledge, this is the first time a graph-based algorithm has been shown to achieve high performance (AUC$>$$0.9$) on new users who have only sent a small number of friend requests.

\end{abstract}

\copyrightyear{2020}
\acmYear{2020}
\acmConference[WWW '20]{Proceedings of The Web Conference 2020}{April 20--24, 2020}{Taipei, Taiwan}
\acmBooktitle{Proceedings of The Web Conference 2020 (WWW '20), April 20--24, 2020, Taipei, Taiwan}
\acmPrice{}
\acmDOI{10.1145/3366423.3380204}
\acmISBN{978-1-4503-7023-3/20/04}



\newcommand{\udi}[1]{\textcolor{red}{#1}}

\newcommand{\sybiledge}{\textsc{SybilEdge}\xspace}

\maketitle
\section{Introduction}
\label{intro}

Online social networks are frequently targeted by malicious actors who create `fake’ or `sybil’ accounts for the purpose of carrying out abuse. Broadly, abuse is conducted in three phases: First, malicious actors create accounts. These accounts then need to establish connections with real users (e.g. by sending friend requests on Facebook). Once they establish sufficient connections, fake accounts can expose their social networks to a variety of malicious activities. 

According to its latest Community Standards Enforcement Report~\citep{Standards2019}, Facebook disabled over $2.2$ billion such accounts in the first quarter of 2019. The vast majority of these accounts were disabled during or within minutes of account creation, and $99.8$\% were disabled before being reported by a Facebook user. Despite these impressive figures, the fraction of such accounts that survives registration-time classifiers and forms connections on Facebook still constituted roughly $5$\% of monthly active users in 2019~\citep{Standards2019}. 

In this paper, we focus on social-graph-based detection of \emph{new} fake accounts that manage to evade registration-time classifiers but have not yet made sufficient connections to perpetrate abuse. We define \emph{new} accounts as those that are less than $7$ days old or have sent fewer than $50$ friend requests.

\begin{figure}[t]
\centering
\includegraphics[height=0.79\textwidth]{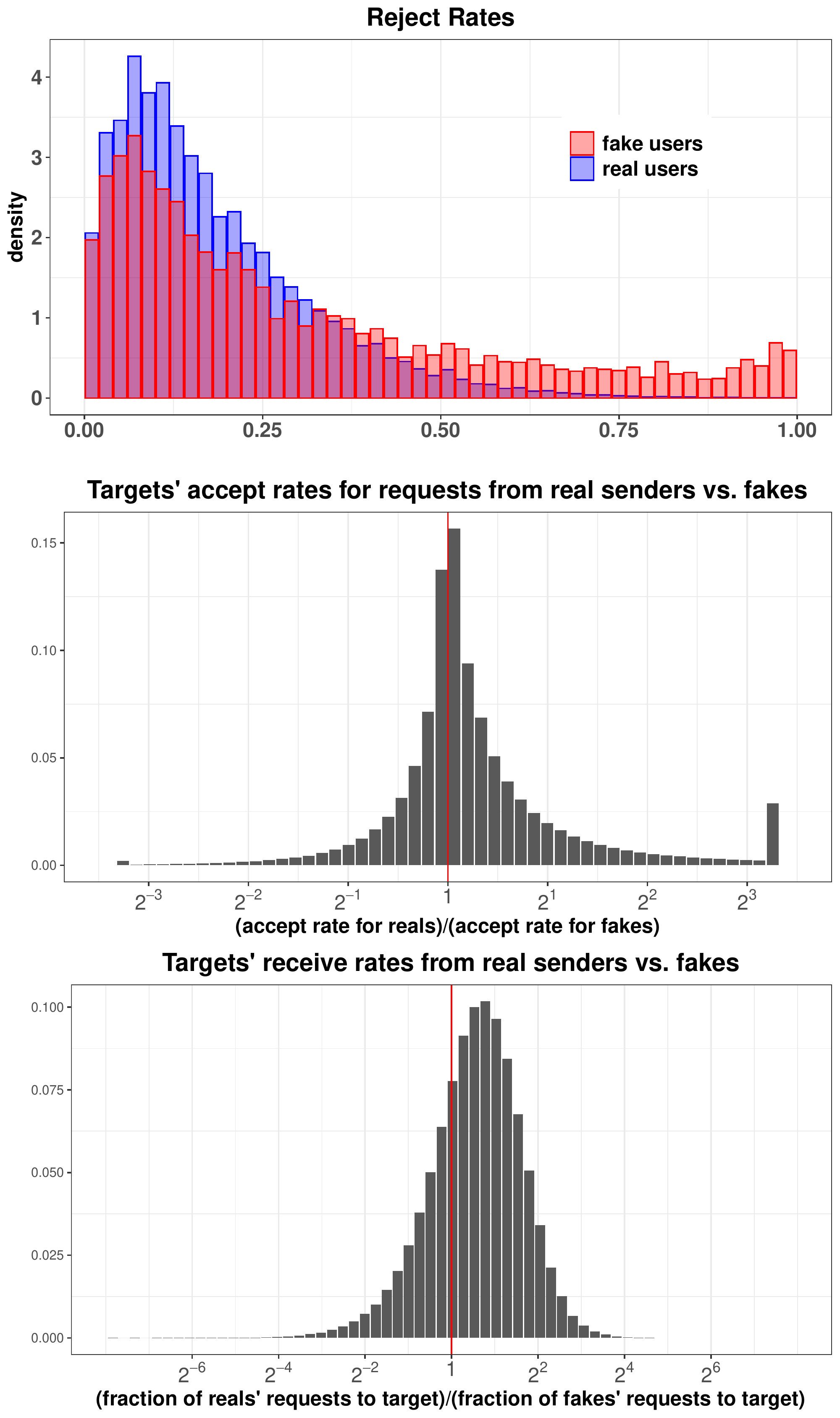}
\caption{\textit{Top:} Distribution of rates at which real and fake Facebook users' friend requests are rejected. \textit{Middle:} Distribution of Facebook users’ ratios of their accept rates for incoming requests from reals and fakes. Users right of the red line at $x$=$1$ are more likely to accept a friend request from a real account than a fake. \textit{Bottom:} Distribution of the ratios of the fraction of reals' requests and the fraction of fakes' requests that target each Facebook user. Mass right of the red line at $x=1$ represents users who receive disproportionately more of real users' requests than fakes' requests. }
\Description[Motivating distributions]{\textit{Top:} Two similar right-skewed distributions. \textit{Middle:} Distribution with much mass centered around zero, long tails, and more mass on the R.H.S. \textit{Bottom:} Quasi-normal distribution centered around x=2.} 
\label{fig:motivation_plots}
\end{figure}

While the general problem of using the social graph to detect fake accounts is well-studied, existing algorithms typically do not apply to new accounts. This is because mainstream graph-based algorithms use a \emph{structural difference} to detect fake accounts---namely, that fake accounts tend to have lower connectivity to real users. When popularized over a decade ago, this approach exhibited a key advantage: it was assumed that online social network companies only knew the true $\{$fake, real$\}$ labels for a handful of users, and this handful was sufficient to seed a graph-based detection algorithm based on this structural difference. Nonetheless, one disadvantage is that real and fake users will only tend to exhibit this structural difference when they have made a reasonable fraction of their connections, so these algorithms tend to exclude new users from effective detection \cite{yang2012sybilrank, boshmaf2015integro, al2017sybilreview, ramalingam2018fakesybilreview}.

However, both the resources available to online social networks and the challenges they face have evolved in the 14 years since the popularization of graph-based algorithms. For example, Facebook now possesses high-confidence \{fake, real\} labels for a majority of its active users---not just the handful assumed by existing graph-based algorithms. It is therefore now possible to use these additional labels to estimate not just structural differences, but also \emph{individual-level differences} in how different users interact with real and fake accounts. Nonetheless, these labels are typically only available for users who have been active for at least several weeks. Thus, it is natural to consider whether today's greater data availability can inform algorithms capable of detecting \emph{new} fake accounts.

Specifically, using Facebook's data on friending activity and known fake accounts, we observe that there are in fact important \emph{individual-level} differences in how fake accounts interact with real users, and how real users react to fake accounts. Observing these differences requires looking beyond aggregate statistics such as a user's overall reject rate for the friend requests she sends. For example, Fig.~\ref{fig:motivation_plots} \emph{top}
shows the distributions of reject rates for fake and real accounts. As the figure shows, many fakes (like many real accounts) either never or rarely have their friend requests rejected.

Disaggregating this data reveals two key differences: First, \emph{for certain users} (but not others), whether a request comes from a 
fake/real account is highly determinative of her decision to accept or reject. Fig.~\ref{fig:motivation_plots} (\emph{middle}) plots the ratios of each Facebook user's rates at which she accepts friend requests from reals and fakes. Mass at $x=1$ represents Facebook users who accept/reject fakes at the same rate as reals, which provides no information about the sender's $\{$fake, real$\}$ label. Mass to the right of $x=1$ represents users who are more likely to accept a request from a real user than one from a fake user by a factor corresponding to the $x$-axis. Thus, an unknown user whose friend request is accepted by such a recipient is more likely to be real. For example, the mass at $x=2^3$ represents users who are $8$ times more likely to accept a request from a real user than a fake. In fact, over $1/3$ of Facebook users are at least $1.5$ times as likely to accept either a real or a fake (i.e., mass outside $(\frac{1}{1.5}, 1.5)$), which provides a strong signal of their senders' labels. Because the tails of this distribution are very wide, we round all users with ratios outside $(\frac{1}{10},10)$ to these bounds in the plot.

Second, we observe a key difference in how some fake accounts select targets for their friend requests \emph{differently} than real accounts. Specifically, certain users tend to be more or less frequently targeted by friend requests from fakes compared to real users, such that sending a request to such a target reveals information about the sender’s label. Fig.~\ref{fig:motivation_plots} \emph{bottom} plots the ratios of the fraction of reals' requests and the fraction of fakes' requests that target each Facebook user. Here, mass at $x=1$ represents users who are equally likely to be selected as the recipient for a fake sender's friend request as for a real sender's friend request. Note that some users (mass to the left of $x=1$) are preferred by fake senders, but many users (mass to the right of $x=1$) are disproportionately likely to be selected by a real sender. Thus, an unknown user who sends a request to such a target to the right of $x=1$ is more likely to be real, and vice versa. Here, $65\%$ of users are at least $1.5$ times as likely to be selected by a real vs. a fake or vice versa (i.e., mass outside $(\frac{1}{1.5}, 1.5)$), which provides a strong signal of their senders' labels.

These two key individual-level differences suggest a new means to detect \emph{new} fake users despite their sparse connections: existing users are unequal in how their acceptances of friend requests reflect information about senders' real/fake labels, and real/fake senders deliberately target their requests to different sets of recipients.

\smallskip
\noindent\textbf{Main contribution.} In this paper we present \textsc{SybilEdge}, an algorithm to identify \emph{new} fake accounts on social networks. \textsc{SybilEdge} returns the probability that each new user is a fake by aggregating over (I) her choices of friend request targets and (II) these targets' corresponding accepts/rejects. 
We show that this algorithm rapidly detects new fake users at scale on the Facebook network and outperforms state-of-the-art benchmark algorithms. We also show that \textsc{SybilEdge} is robust to label noise in the training data, to greater prevalence of fake accounts in the network, and to several different ways fakes can select targets for their friend requests. To our knowledge, this is the first time a graph-based algorithm has been shown to achieve high performance (AUC$>$$0.9$) on new users who have sent only a small number of friend requests.

\smallskip
\noindent\textbf{Technical overview}.
 \textsc{SybilEdge} classifies new users by combining three key components: First, \textsc{SybilEdge} estimates whether a new user is a fake by aggregating over her \emph{choices of friend request targets}, giving more weight to targets to the extent they are preferred by other fakes vs. real users. Second, \textsc{SybilEdge} aggregates over these \emph{targets’ responses} (accept/reject) to the user’s friend requests, giving more weight to targets to the extent they respond differently to fakes versus real users. Finally, during these aggregations, \textsc{SybilEdge} gives more weight to choices of targets and their responses when we are more \emph{confident} that they distinguish fakes from real users. Together, these three components give \textsc{SybilEdge} a natural means to elicit the information about a new user’s $\{$fake, real$\}$ identity from each of her friendship edges.

\subsection{Related work}
\label{ssec:related}

A variety of work has proposed graph-based algorithms to detect fake accounts \citep{yu2006sybilguard, yu2008sybillimit, yang2012sybilrank, gong2014sybilbelief, danezis2009sybilinfer, jia2017random, boshmaf2015integro, wang2017gang, wang2017sybilscar, al2017sybilreview, xue2013votetrust, wang2018sybilblind, yang2012analyzing, mohaisen2011keep, qi2018detecting, subrahmanian2016darpa, wang2018structure, yang2014uncovering, alvisi2013sok, mislove2010you, cao2015combating, zhang2015truetop, fu2017robust}. Mainstream graph-based algorithms typically proceed from the \emph{homophily assumption}, which assumes that a pair of connected users shares the same $\{$fake, real$\}$ label with high probability, such that fakes tend to be poorly connected to real users overall \cite{yu2006sybilguard, yu2008sybillimit, yang2012sybilrank, gong2014sybilbelief, wang2017sybilscar, al2017sybilreview}. Based on this assumption, a variety of graph-based algorithms attempt to propagate trust out from a small known set of trusted real users to unknown ones based on their connectivity to the known set.

More specifically, these algorithms typically propagate trust outwards via either random walks or Markov random fields (i.e. loopy belief propagation methods). Random walk based methods proceed on the basis of the assumption that unknown real users will be reachable in relatively few hops from the known set of real users, whereas reaching fake accounts requires additional hops on average. These algorithms therefore typically proceed via a series of short random walks on the network to partition nodes into real and fake sets on this basis. Random walk based methods include the seminal \textsc{SybilGuard} algorithm ~\cite{yu2006sybilguard}, as well as \textsc{SybilLimit} ~\cite{yu2008sybillimit}, \textsc{SybilInfer} ~\cite{danezis2009sybilinfer}, \textsc{SybilWalk} ~\cite{jia2017random}, \textsc{Integro} ~\cite{boshmaf2015integro}, and \textsc{SybilRank} ~\cite{yang2012sybilrank}. 
Importantly, while random walk based approaches require either a known set of real users or a known set of fakes, they cannot leverage both at the same time. They are also considered less robust to misclassification (i.e. label noise) in the set of known users ~\citep{wang2017sybilscar}.

In contrast to random walk based methods, loopy belief propagation methods take a probabilistic view. These methods use Markov random fields to capture network structure and define a joint probability distribution over each node's label, which is iteratively updated to propagate labels from known fake or real nodes to unknown ones. Algorithms of this type include the seminal \textsc{SybilBelief} ~\cite{gong2014sybilbelief}, \textsc{Sybilfuse} ~\cite{gao2018sybilfuse}, and \textsc{GANG} ~\cite{wang2017gang}. Such algorithms are able to incorporate information about both known real and known fake nodes, and they are also robust to some noise in this set of known labels. Recently, Wang et al. proposed a hybrid algorithm, \textsc{SybilSCAR} \citep{wang2017sybilscar}, based on this approach. \textsc{SybilSCAR} iteratively propagates probabilistic estimates of unknown nodes' labels based on a known set of users of each type.

Importantly, both types of algorithm require that all users have had sufficient `stabilization time' to make the majority of their connections such that they will exhibit the homophily assumption~\citep{yang2012sybilrank, boshmaf2015integro, al2017sybilreview, ramalingam2018fakesybilreview}. Due to these requirements, evaluations of fake detection algorithms have often excluded users with less than e.g., $1$ to $6$ months of tenure on the social network~\citep{yang2012sybilrank, boshmaf2015integro}, which provides an ample `grace period' for fake accounts to perpetrate abuse.

One partial exception is \textsc{VoteTrust} \cite{xue2013votetrust}. \textsc{VoteTrust} assumes that a majority of users (including those with known real labels) will be long-tenured, but this long-tenured set can be leveraged to classify a new user. For \textsc{VoteTrust}, however, this advantage comes at a cost: \textsc{VoteTrust} requires the additional dataset of (ideally) all historical friendship requests in the history of the social network, or at very least, sufficient historical requests such that the directed graph of requests is connected~\citep{xue2013votetrust}. We note that data on old friendship requests is typically not among the datasets considered to be readily accessible for analysis in the current generation of online social networks.

 The homophily assumption may also cause these algorithms to misclassify when some `successful' fake accounts succeed in connecting to many real accounts. Recent research ~\citep{freeman2017can, ghosh2012understanding} suggests that this phenomenon is relatively prevalent on social networks. For the same reason, the homophily assumption renders these algorithms vulnerable to \emph{sampling attacks} whereby a malicious user defeats these algorithms by instructing some of her fake accounts to send many friend requests to real users (knowing that many of these accounts may be detected), then instructing her remaining fake accounts to send requests only to the subset of real users who were willing to accept requests from fakes. By generating fake users who are densely connected to real accounts, the attacker may succeed in convincing an algorithm that fake users are real \cite{xue2013votetrust,cao2013sybilfence}.


\medskip

\noindent\textbf{Paper organization.}
We present the \textsc{SybilEdge} algorithm in Section~\ref{sec:Model}. We evaluate the performance of \textsc{SybilEdge} on the Facebook network in Section \ref{sec:evaluations}. We study \textsc{SybilEdge}'s robustness to label noise in Section \ref{sec:labelnoise} and its robustness to the prevalence of fake accounts in Section \ref{sec:Experiments}. We conclude the paper in Section \ref{sec:conclusion}.


\section{The SybilEdge algorithm}
\label{sec:Model}
In this section we derive the \textsc{SybilEdge} ($E$xpert $D$ecision $G$iven $E$dges) algorithm and its three key components: \emph{target selection, target response,} and \emph{confidence weighting}.

\smallskip
\noindent \textbf{Preliminaries.} Our goal is to determine the posterior probability $p_i$ that a new user $i$ is fake as a function of the set $T_i$ of targets (friend request recipients) to whom she sends friend requests and their respective responses. Let $\delta_i \in \{S,B\}$ represent user $i$'s label as a fake/sybil ($S$) or real/benign ($B$) account---that is, the label we want to learn. Let $x_{ij} \in\{0,1\}$ denote target $j$'s response to $i$'s friend request (i.e. accept or reject), where $x_{ij}=1$ denotes that $j$ accepted $i$'s request, and let $X_i \in \{0,1\}^{|T_i|}$ denote the binary vector of all responses $x_{ij}$ to $i$'s requests from her set of targets $T_i$.

We denote by $r_j^S$ (and $r_j^B$) an arbitrary fake (and real) sender's probability of choosing user $j$ as the target when she sends her first friend $r$equest.  We denote by $R_i^S$ the vector of probabilities $r_j^S$ for all of $i$'s targets $T_i$, and by $R_i^B$ the corresponding vector of probabilities $r_j^B$ for all of $i$'s targets. We denote the probability that $j$ $a$ccepts a request from a fake or a real sender as $a_j^S$ and $a_j^B$, respectively. Similarly, we denote by $A_i^S$ the vector of accept probabilities $a_j^S$ for all of $i$'s targets, and by $A_i^B$ the vector of accept probabilities $a_j^B$ for $i$'s targets.

Finally, suppose we know a $l$abelled set $L = L^S \cup L^B$ of known fake and real users, where $L^S$ is the set of known fakes and $L^B$ is the set of known real users, and suppose we have prior knowledge $\pi_i$ of $i$'s label.

\begin{table}
	\begin{tabular}{cp{6.5cm}}
		\textbf{Notation} & \textbf{Description} \\
		\hline
		$i$ & The user whose label we infer;\\
		$\delta_i \in \{S,B\}$ & User $i$'s label:  fake ($S$) i.e. sybil, or real ($B$) i.e. benign;\\
		$\pi_i$ & Prior on user $i$'s probability of being a fake;\\
		$p_i$ & Posterior probability that $i$ is a fake;\\
		$T_i$ & The set of targets $i$ sends friend requests to;\\
		$r_{j}^S$, $r_{j}^B$ & Probabilities that user $j$ is the target of an arbitrary fake and real sender's first friend request, resp.;\\
		$R_i^S$, $R_i^B$  & The vectors of probabilities $r_j^S$ and $r_j^B$, respectively for all targets $j$ whom $i$ sends requests to;\\
        $x_{ij} \in\{0,1\}$ & Target $j$'s response to $i$'s request ($1 =$ accept);\\
        $X_i$ & Obs. responses $[x_{i1}, \ldots, x_{i|T_i|}]$ of all $i$'s targets;\\
		$a_j^S$, $a_j^B$ & Probabilities that target $j$ accepts a friend request from a fake and from a real sender, respectively;\\
		$A_i^S$, $A_i^B$ & Vectors of probabilities $a_j^S$ and $a_j^B$, respectively for all targets $j$ whom $i$ sends requests to;\\
		$L, \ L^S, L^B$ & Sets of known (users, fakes, reals);\\
		$\rho_j, \rho_j^S, \rho_j^B$ & Counts of known (users, fakes, reals) who sent requests to $j$; \\
		$f_j, f_j^S, f_j^B$ & Counts of known  (users, fakes, reals) whose requests $j$ accepted;\\
		$\sigma_j, \phi_j$ & Priors on target $j$'s quality as a classifier of users who send requests to $j$ and are accepted by $j$, resp.\\
		\hline
	\end{tabular}
\caption{Notation used in \textsc{SybilEdge}}
\Description[Notation used in \textsc{SybilEdge}]{} 
\label{tab:notations}
\end{table}

\subsection{Component I: a user’s selection of targets}
Here, we derive the first component of \textsc{SybilEdge}, which updates our estimate of whether user $i$ is a fake based on whether she selects targets for her friend requests that are preferred by known fake versus known real senders. Specifically, we model that each new user selects a target for her first friend request via a draw from a multinomial distribution corresponding to her $\{$fake, real$\}$ label: Fake users select each target $j$ with probability $r_j^S$, but real users select $j$ with probability $r_j^B$.  
We can then estimate the posterior probability that a sender $i$ is fake based on the relative probabilities that a fake/real user would have selected $i$'s set $T_i$ of targets:

\begingroup
\fontsize{8pt}{10.5pt}\selectfont
\begin{align}
Pr\big[\text{$\delta_i$=$S$}|T_i, R_i^S, R_i^B,\pi_i]=\frac{\pi_i Pr\big[T_i| \text{$\delta_i$=$S$},  R_i^S\big]}{\pi_i Pr\big[T_i| \text{$\delta_i$=$S$},  R_i^S\big] + (1\text{-}\pi_i)Pr\big[T_i| \text{$\delta_i$=$B$}, R_i^B\big]  }\label{basemodel_derive}
\end{align}
\endgroup


\noindent Where $R_i^S$ and $R_i^B$ denote the vector of all probabilities $r_j^S$ and $r_j^B$, respectively, for the targets $j\in T_i$ to whom user $i$ sends requests.

We assume that conditional on the sender's label $\delta_i$, the relative probability that a sender selects any target $j$ is conditionally independent\footnote{This is a standard assumption (see e.g. \citep{raykar2010learning}). While not true in general (e.g. some targets are more popular), this assumption is advantageous as it may limit the effect any one observation has on model predictions, rendering it more adversarially robust.} of everything else, and that the count $|T_i|$ of friend requests the sender sends is independent of her label.\footnote{The assumption that a user's count $|T_i|$ of friend requests is independent of her label is advantageous because is allows \textsc{SybilEdge} to apply equally to accounts that are e.g. $1$ and $7$ days old---that is, accounts that have sent fewer/more friend requests.} Technically, as a user $i$ sends more friend requests, she reduces the remaining set of possible targets for her next friend request, making each of them slightly more probable for the next request. However, because the network is very large compared to any user's number of friend requests, sampling targets with replacement is a very good approximation of sampling without replacement.\footnote{Absent this approximation, we would re-normalize $r_j^S$ and $r_j^B$ after each subsequent request, so e.g. the numerator in eq. \ref{selectioncomponent} would become $\prod_{j \in T_i}  r_j^S / \textstyle (\sum_{k{\in L\backslash T_i[:j]}}r_k^S)$, where $T_i[:j]$ denotes the targets to whom she sent requests before $j$.}

\smallskip
\noindent Thus, we can then compute this \textbf{target selection component} via:
\begin{align}
p_i = \frac{\overbrace{\pi_i}^{\text{prior}} \displaystyle \prod_{j \in T_i}  r_j^S  }{\pi_i \displaystyle \prod_{j \in T_i} r_j^S + (1-\pi_i)\displaystyle \prod_{j \in T_i} r_j^B}
\label{selectioncomponent}
\end{align}

Here, the numerator is the joint probability of sender $i$'s selections of friend request targets $T_i$ given these targets’ probabilities at which they are selected by \emph{fake} accounts. The denominator then gives the total probability of $i$’s selections of targets, which we compute by adding the probability of these selections given that the sender $i$ was \emph{fake} plus the probability that they occurred given that sender $i$ was \emph{real}. Therefore, the entire expression gives the \emph{relative} probability that $i$ is fake given her selections of targets, scaled by $\pi_i$, the prior probability that $i$ is fake (for example, we might set this the overall fraction of fake accounts at Facebook).

The key intuition is that eq. \ref{selectioncomponent} only updates our posterior estimate that $i$ is fake to the extent her targets are selected by fake and real users at different rates (i.e. to the extent that $i$ sends requests to targets who are further from $x$=$1$ in Fig. \ref{fig:motivation_plots}, \emph{bottom}). In section \ref{confidence}, we show how to estimate targets' selection rates $r_j^S$ and $r_j^B$.

\subsection{Component II: targets’ responses}
Here, we derive the second component of \textsc{SybilEdge}, which updates our estimate of whether user $i$ is fake based on her targets' \emph{responses} to her friend requests. Suppose (unlike above) that a target is equally likely to receive a friend request from an arbitrary real or fake account, such that receiving a friend request from a user reveals no information about that user’s label.\footnote{In this case, $r_j^S=r_j^B,\forall j$, so eq. \ref{selectioncomponent} factors to the prior $\pi_i$.} However, suppose we observe the targets' responses (acceptances/rejections) $x_{ij}$ of $i$'s friend requests, and targets may \emph{accept} fake senders' requests at different rates than real senders' requests. If we know each target's probabilities $a_j^S$ and $a_j^B$ of accepting a request from a fake sender and from a real sender, respectively, then we can use the sequence of observed responses $X_i$ to each of user $i$'s friend requests to estimate the probability that she is fake. Denote by $A_i^S$ and $A_i^B$ the vectors of probabilities $a_j^S$ and $a_j^B$, respectively, for all targets $j\in T_i$ to whom $i$ sends requests. Assume that conditional on the sender's label $\delta_i$, targets' responses are conditionally independent of everything else. We estimate the probability $i$ is fake via:

\begingroup
\fontsize{8pt}{10.5pt}\selectfont
\begin{align}
Pr[\text{$\delta_i$=$S$}|X_i, A_i^S, A_i^B, \pi_i]=\frac{\pi_i Pr\big[X_i| \text{$\delta_i$=$S$},  A_i^S\big]}{\pi_i Pr\big[X_i| \text{$\delta_i$=$S$},  A_i^S\big] + (1\text{-}\pi_i)Pr\big[X_i| \text{$\delta_i$=$B$}, A_i^B\big]  }\label{responsecomponent_deriv}
\end{align}  
\endgroup


We now show how to compute this probability. Because a target may accept or reject a request, we first simplify notation by defining a function $\euscr{A}(x_{ij}, \delta_i)$ that takes two inputs: target $j$'s accept or reject $x_{ij}$ of $i$'s friend request, and the indicator $\delta_i$ of whether the source $i$ is fake or real. $\euscr{A}(x_{ij}, \delta_i)$ returns the probability that target $j$ accepts $i$ conditional on her $\{$fake, real$\}$ label if we observe that $i$'s friend request was accepted by $j$, or the complement of this probability if we observe that $i$ was rejected by $j$:

\[
\euscr{A}(x_{ij}, \delta_i)= 
     \begin{cases}
      a_j^\delta &if \ x_{ij}=1;\\
      1-a_j^\delta &if \ x_{ij}=0
     \end{cases}\\\\
\]
\bigskip
\newline
Now we can compute this \textbf{target response component} via:

\begin{align}
p_i=\frac{\overbrace{\pi_i}^{\text{prior}}
 \displaystyle \prod_{j \in T_i}  \euscr{A}(x_{ij}, \text{$\delta_i$=$S$})}{\pi_i \displaystyle \prod_{j \in T_i} \euscr{A}(x_{ij}, \text{$\delta_i$=$S$}) + (1-\pi_i)\displaystyle \prod_{j \in T_i} \euscr{A}(x_{ij}, \text{$\delta_i$=$B$})} \label{responsecomponent}
\end{align}

Here, the product in the numerator is the probability of observing sender $i$'s accepts and rejects conditional on her targets using the probabilities at which they accept and reject \emph{fake} accounts. The denominator then gives the total probability of observing these accepts and rejects. 
At a high level, the entire expression captures the question `did source $i$'s accepts/rejects appear to be due to her targets treating her as they treated fakes or as they treated reals?'.

The key aspect to note is that eq. \ref{responsecomponent} only updates the posterior estimate that $i$ is fake to the extent her targets respond differently to requests from fakes vs. reals (i.e. to the extent that $i$'s targets are further from $x$=$1$ in Fig. \ref{fig:motivation_plots}, \emph{middle}). In section \ref{confidence}, we show how to estimate targets' accept rates $a_j^S$ and $a_j^B$ for each class of senders.

\subsection{The SybilEdge equation}
Here we show how to compute the key equation in the \textsc{SybilEdge} algorithm, which combines these target selection and target response components to aggregate the information about a user’s $\{$fake, real$\}$ label contained in each of her friendship edges. Specifically, we say that the probability of observing each of $i$'s accepted or rejected edges can be decomposed as (I) the probability that $i$ would select the edge's target conditional on $i$'s $\{$fake, real$\}$ label, and (II) the target's response conditional on $i$'s selection of the target and $i$'s label. We thus determine the posterior probability $i$ is a fake by aggregating over $i$'s edges via \textbf{the \textsc{SybilEdge}} \textbf{equation}:

\begin{align}
p_i=\frac{\overbrace{\pi_i}^{\text{prior}} \displaystyle \prod_{j \in T_i} \overbrace{ \euscr{A}(x_{ij}, \text{$\delta_i$=$S$})}^{\text{target response}} \cdot \overbrace{r_j^S}^{\text{ selection}}}{\pi_i \displaystyle \prod_{j \in T_i} \euscr{A}(x_{ij}, \text{$\delta_i$=$S$}) \cdot r_j^S + (\text{$1$-$\pi_i$})\displaystyle \prod_{j \in T_i} \euscr{A}(x_{ij}, \text{$\delta_i$=$B$}) \cdot r_j^B} \label{sybiledgeeqn}
\end{align}

Here, the products in the numerator give us the joint probability that (I) $i$ selects the set of targets to whom she sends friend requests \emph{as a fake user would select targets}; and (II) these targets respond with the accepts and rejects we observe given that they treat $i$ as a fake when accepting/rejecting her. The products in the denominator then give the total probability that $i$ selects these targets and they respond with the accepts/rejects we observe. The \textsc{SybilEdge} equation therefore gives us the relative probability that the $i$'s set of requests, accepts, and rejects are those of a fake user.

The \textsc{SybilEdge} equation thus captures our intuitions that a user $i$ is more likely to be a fake to the extent that she selects targets who are preferred by fakes (for whom $r_j^S>r_j^B$), and also to the extent her targets respond differently to her requests than they usually respond to requests from reals (for whom $\euscr{A}(x_{ij}, \text{$\delta_i$=$S$}) > \euscr{A}(x_{ij}, \text{$\delta_i$=$B$)}$).

\subsection{Component III: weighting target confidence}
\label{confidence}
The discussion above assumes we know the true probabilities at which fakes and reals each select each target ($r_j^S$ and $r_j^B$), and the probabilities at which each target accepts a request from either class ($a_j^S$ and $a_j^B$). In practice, we must estimate these parameters from observed social graph data. Therefore, we introduce the final component of the \textsc{SybilEdge} algorithm: \textsc{SybilEdge} gives more weight to selections of targets and targets’ responses not only as a function of the magnitude of the difference of targets’ request and accept probabilities for fakes vs. real users (as above), but also as a function of our \emph{confidence} in these differences.

\textsc{SybilEdge} accomplishes this confidence weighting as follows.
 First, consider how to compute $a_j^S$, $a_j^B$, that is, each target's probability of accepting a friend request from an arbitrary fake or real user. Suppose we know a set $L^S$ of existing fakes and a set $L^B$ of existing real users. The maximum-likelihood estimate of $a_j^S$ is just target $j$'s count of accepts of the requests she received from known fakes divided by the total count of these requests. However, if we used this approach for all targets, then the \textsc{SybilEdge} equation would give equal weight to a target who responded to only a few requests (i.e. a target whose accept rates we know with low confidence) and a target had responded to thousands of requests (whose accept rates we know with high confidence).

Therefore, we instead use estimators for these rates that, \emph{in the absence of data to the contrary}, shrink $a_j^S$ and $a_j^B$ towards each other. This is because in the case where $a_j^S = a_j^B$, we say target $j$ is equally likely to accept $i$'s friend request regardless of whether $i$ is a fake or a real, so the fact that $j$ accepts $i$ does $not$ update $i$'s probability of being a fake according to the \emph{target response} component of the \textsc{SybilEdge} equation above.

Specifically, let $f_j$ denote the count of $j$'s \emph{acceptances} of friend requests from users with known labels, and let $f_j^S$ and $f_j^B$ denote the counts accepted from known fakes and known reals, respectively. Let $\rho_j$ denote the count of all friend requests that known users sent to target $j$, and let $\rho_j^S$ and $\rho_j^B$ denote $j$'s count of friend requests from just known fake senders and just known real senders, respectively. We use estimators that \textbf{reweight target accept rates} based on our confidence via:
\begin{align}
\hat{a}_j^{overall} = \frac{f_j}{\rho_j}; \ \hat{a}_j^S = \frac{f_j^S + \phi_j \cdot \hat{a}_j^{overall}}{\rho_j^S + \phi_j}; \ \hat{a}_j^B = \frac{f_j^B + \phi_j \cdot \hat{a}_j^{overall}}{\rho_j^B + \phi_j}
\label{targetrate}
\end{align}

\noindent Where $\phi_j:\phi_j\ge0$ is a `confidence' prior on target $j$ for the \emph{target response} component of \textsc{SybilEdge}. Setting $\phi_j=0, \forall j$ recovers the maximum likelihood estimators for $\hat{a}_j^S$ and $\hat{a}_j^B$, which compel the \textsc{SybilEdge} equation to place equal weight on targets for whom we have observed more or less acceptance data.\footnote{There is a mathematical equivalence between these estimators and the Beta conjugate model in Bayesian inference.} In contrast, as we increase $\phi_j$, we shrink $\hat{a}_j^S$ and $\hat{a}_j^B$ together \emph{to a degree that is inversely proportional to the count of friend requests $j$ responded to}, which compels \textsc{SybilEdge} to place less weight on targets who have only accepted/rejected a small number of reals or fakes in the past. In this case, the \textsc{SybilEdge} equation will tend to learn only from targets whom we have \emph{repeatedly} observed accepting reals at a different rate than fakes (i.e. targets whose acceptance rates are known with high confidence). Similarly, by increasing $\phi_j$ for a particular target $j$ but not others, we can selectively downweight the influence of target $j$'s accepts/rejects on the model's predictions, which may be advantageous if we suspect target $j$ of being a malicious or adversarial user.

\textsc{SybilEdge} uses a similar approach to place more weight in the \emph{target selection} component on targets when we are more confident (i.e. have observed more data) about how they are selected by fakes vs. reals. Similarly to above, we could imagine maximum likelihood estimators for the probability $r_j^S$ (or $r_j^B$) that a fake (or real) user will send her first friend request to target $j$ by computing target $j$'s count of requests received from known fakes (or reals) divided by the count of all requests sent by known fakes (or reals). 
But, as above, this approach would cause \textsc{SybilEdge} to give equal weight to a target who received only a few requests (i.e. a target whose rates we know with low confidence) and a target who received thousands of requests (whose rates we know with high confidence).

Therefore, we instead reweight target selection rates based on our confidence. Specifically, let $\rho_L$ denote the count of all friend requests sent by known users, and let $\rho_{L^S}$ and $\rho_{L^B}$ denote the counts sent by known fakes and reals, respectively. Instead of the maximum likelihood estimators described above, we use the following to \textbf{reweight target selection rates:}
 
\begin{align}
\hat{r}_j^{overall}=\frac{\rho_j}{\rho_{L}};
\ \hat{r}_j^S=\frac{\rho_j^S + \sigma_j \cdot \hat{r}_j^{overall}}{\rho_{L^S} + \sigma_j}; \ \hat{r}_j^B=\frac{\rho_j^B + \sigma_j \cdot \hat{r}_j^{overall}}{\rho_{L^B} + \sigma_j}
\label{sendrate}
\end{align}

Here, $\sigma_j:\sigma_j\ge0$ is our `confidence' prior on target $j$ for the \emph{target selection} component of the \textsc{SybilEdge} equation: if we set $\sigma_j=0, \forall j$, the \textsc{SybilEdge} equation places \emph{equal} weight on friend requests sent to targets for whom we have observed more/less data; increasing $\sigma_j>0, \forall j$ causes the \textsc{SybilEdge} equation to place \emph{more weight} on targets for whom we have observed more data. More specifically $\sigma_j=0$ recovers the maximum likelihood estimators for $r_j^S$ and $r_j^B$, whereas increasing $\sigma_j>0$ shrinks $\hat{r}_j^S$ and $\hat{r}_j^B$ towards each other \emph{to a degree that is inversely proportional to the overall count of friend requests $j$ received from fakes or reals}. This in turn causes the \textsc{SybilEdge} equation to place less weight on learning from targets for whom we have observed fewer friend requests (recall that the \emph{target selection} component only updates the probability that a user is fake to the extent that fake users send requests to her targets at different rates than real users). Similarly, by setting $\sigma_j$ higher for a particular target $j$ compared to others, we downweight the influence of the selection of target $j$ compared to other targets in the \textsc{SybilEdge} equation.

\subsection{The SybilEdge algorithm}
These \emph{target selection, target response}, and \emph{confidence weighting} components form the \textsc{SybilEdge} algorithm:

\begin{algorithm}[H]
\caption{\textsc{SybilEdge}}
\begin{algorithmic}
    	\INPUT $G_{requests}(V,E), G_{accepts}(V,E'), L,\pi, \sigma, \phi$
    	\STATE \textbf{for} known user $j \in L$
    	\STATE \ \ \ \ compute weighted request rates $\hat{r}_j^S$ and $\hat{r}_j^B$ per \emph{eq.} \ref{sendrate}\\
    	\STATE \ \ \ \ compute weighted accept rates $\hat{a}_j^S$ and $\hat{a}_j^B$ per \emph{eq.} \ref{targetrate}\\  
        \STATE \textbf{for} new user $i \in V\backslash L$
        \STATE \ \ \ \ compute $p_i$ per \emph{eq.} \ref{sybiledgeeqn}
    	\STATE \textbf{return} $p_i$ for all $i$
  \end{algorithmic}
  \label{alg:full}
\end{algorithm}

\subsection{Choosing tuning parameters $\phi$ and $\sigma$}
\label{ssec:choosing_phisigma}
A key property of tuning parameters $\phi_j$ and $\sigma_j$ is that, by increasing one relative to the other, we can tune \textsc{SybilEdge} to place more emphasis on learning from the set of targets a user chooses to send requests to relative to learning from whether those targets accept or reject. Specifically, as we increase $\sigma_j\rightarrow \infty, \forall j$, \textsc{SybilEdge} sets $\hat{r}_j^S \approx \hat{r}_j^B,\ \forall j$. The algorithm then ceases to update its estimate of $i$'s label based on the set of targets $i$ chooses, and we recover the \emph{target response component} from the full \textsc{SybilEdge} algorithm. This in turn makes \textsc{SybilEdge} more robust to attack, as a fake user cannot `appear real' by sending requests to recipients who typically are not targeted by fakes. However, this robustness comes at a cost in terms of \textsc{SybilEdge}'s recall. Consider, for example, that when all $\sigma_j$ are large, we will be less likely to detect a fake account that sends requests to targets who receive proportionally many requests from fakes, but who accept fakes at the same rate they accept reals.

\subsection{SybilEdge properties}
\label{ssec:adv_robust}

In addition to its strong performance on real and simulated Facebook data, \textsc{SybilEdge} exhibits six advantageous properties:

\smallskip
\noindent\textbf{Rapid classification of new users.} Previous methods typically require a lengthy `stabilization period' before a new account can be classified, and are generally \emph{less} likely to correctly classify a fake account that succeeds in making many friends with real users (even if those users are not discriminating). In contrast, \textsc{SybilEdge} becomes increasingly likely to identify a fake as she (1) sends more friend requests; (2) sends requests to more discriminating targets who accept fakes at a different rate than they accept reals (increasing the difference between $a^S_j$ and $a^B_j$ for $i$'s targets); (3) sends requests to targets who are more often victimized by requests from fake accounts (increasing the difference between $r^S_j$ and $r^B_j$); and (4) sends requests to targets who are more active users (for whom we have greater confidence in $a^S_j$ and $a^B_j$). 

\smallskip
\noindent \textbf{Robustness to sampling attacks.} A key property of \textsc{SybilEdge} is that targets only carry weight in the model to the extent that they receive and accept friend requests from real and fake users at different rates. Thus, a fake account cannot improve the \textsc{SybilEdge}'s estimate of her probability of being fake even if she identifies and connects to many real users who accept e.g. \emph{all} requests indiscriminately. Note that an indiscriminately accepting target has $a_j^S=a_j^B$, which causes the target's accept or reject to appear on both the numerator and denominator of the \emph{target response} component of the \textsc{SybilEdge} equation. This target's response then factors out and has no effect on our posterior estimate $p_i$ of $i$'s label.

\smallskip
\noindent\textbf{Low complexity.} \label{ssec:complexity} \textsc{SybilEdge} has complexity $O(|E|)$ where $E$ is the set of friend requests. Because social networks are typically sparse \cite{mislove2007measurement, gong2014sybilbelief}, we have $O(|E|)=O(|V|)$. This compares favorably to state-of-the-art algorithms such as \textsc{SybilBelief} and \textsc{SybilSCAR}, which require $O(k|E'|)$, where $k$ is the number of iterations (at least $O(\log(|V|)$) and $E'$ is the set of accepted friend requests~\citep{wang2017sybilscar,gong2014sybilbelief}. 
     
\smallskip
\noindent\textbf{Interpretability.}  Unlike mainstream sybil detection algorithms, \textsc{SybilEdge} is interpretable. For example, \textsc{SybilEdge} might classify a user as fake with high probability \emph{because} her friend requests were rejected by specific users who tend to accept all requests from real users and reject those from fakes, and \emph{because} she also sent requests to other users who are preferred targets of fakes. Such interpretability enables researchers to audit the model's classifications---an important precondition for disabling fake accounts.
    
\smallskip
\noindent\textbf{Probabilistically labelled training data.} \textsc{SybilEdge} accepts probabilistically labelled training data rather than binary $\{fake, real\}$ labelled data if desired. For example, an acceptance of a request from a user that data suggests is fake with probability $0.25$ can be input as an acceptance of $0.25$ fake users and $0.75$ real users. 

\smallskip
\noindent\textbf{Robustness to label noise in the training data.} In Section \ref{sec:labelnoise} below, we show that \textsc{SybilEdge} is robust to the presence of misclassified users in the training dataset $L$ of known fake/real users.


\section{Evaluations}
\label{sec:evaluations}
Our goal in this section is to show that \textsc{SybilEdge} achieves high performance (AUC$>$$0.9$) on new users at scale on the Facebook network, and that it significantly outperforms state-of-the-art benchmark algorithms. In subsequent sections we also show that \textsc{SybilEdge} is robust (i) to label noise in the training data, (ii) to greater prevalence of fake accounts in the network, and (iii) to several different ways fakes can select targets for their friend requests.

\subsection{Evaluation on the Facebook network}
\label{sec:real_results}

\begin{figure}[t]
\centering
    \includegraphics[width=0.64\linewidth]{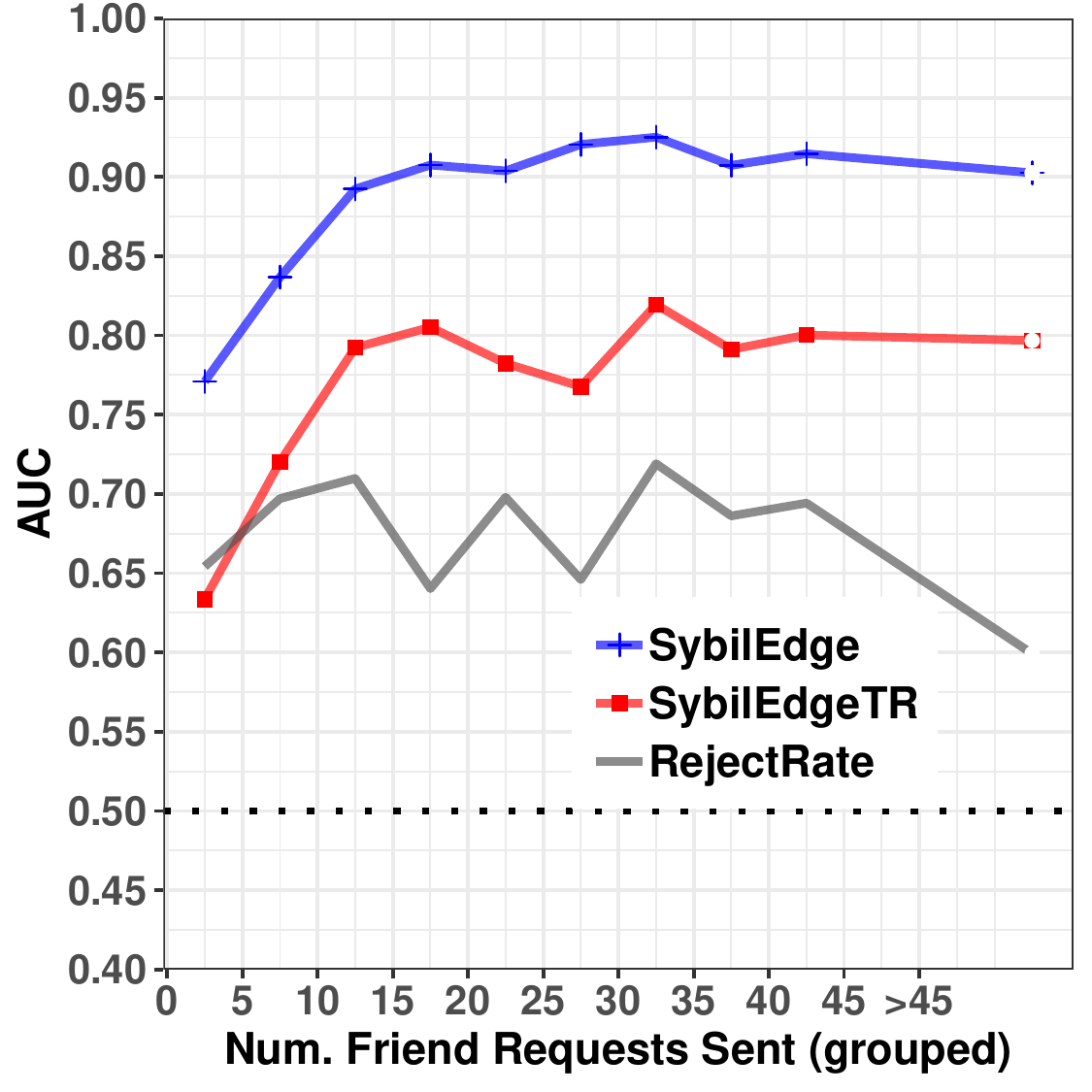}
 \caption{Performance of \textsc{SybilEdge} \ (blue) and \textsc{SybilEdgeTR} (red) on new global Facebook users partitioned by the number of friend requests they sent: $[[0,5], [6,10],\ldots,[41,45],[46,\infty]]$.}
 \Description[\textsc{SybilEdge} and \textsc{SybilEdgeTR} on the global Facebook network]{Plot of AUC vs num. friend requests sent. \textsc{SybilEdge} converges to $AUC>0.91$ after $>15$ requests, and \textsc{SybilEdgeTR} converges to $AUC>0.8$.} 
 \label{FBAUC}
\end{figure}

We implemented \textsc{SybilEdge} at scale at Facebook, and we ran it in an offline evaluation setting on the global Facebook network. Specifically, we trained \textsc{SybilEdge} using just a three-month period of historical friending data from the last year. To train the model, we also used the historical set of real/fake labels from Facebook's internal fake classifiers from these three months. These labels include a highly calibrated real/fake label for all accounts that are $>$$30$ days old, which provided a label for all users in our three months of training data. We then tested \textsc{SybilEdge} by attempting to classify new users who joined Facebook anytime in the week immediately following these three months \emph{using only this one week of data on their friending activity}. That is, we test \textsc{SybilEdge}'s ability to detect new accounts who are each between 0 and 7 days old.\footnote{To ensure fairness in this evaluation, for all new users $i$ we set a prior $\pi_i$ equal to the overall fraction of fakes among new Facebook users, and for all known targets $j$ we set confidence priors $\sigma_j=\phi_j=$ \emph{const.}} Because significant additional time has now passed since these users joined Facebook, they have since been labeled via our same set of fake classifiers. We compare \textsc{SybilEdge}'s output to these known labels.

\smallskip
\noindent\textbf{Comparison metrics and benchmarks.} Due to imbalance in the classes of fakes and real nodes (guessing `all real' yields 95\% accuracy), we adopt the standard approach and use ROC AUC to measure \textsc{SybilEdge's} performance~\citep{wang2017sybilscar, boshmaf2015thwarting}. Recall that an AUC of $0.5$ means a classifier is no better than random on the test set.

 For comparison, we also include two benchmarks: \textsc{RejectRate} and \textsc{SybilEdgeTR} (Section \ref{ssec:realdata_estonia} below adds additional benchmarks).
 
\smallskip
\noindent\textbf{\textsc{RejectRate}.} \textsc{RejectRate} just computes the AUC of each new user's fraction of sent friend requests that are rejected by her targets. 

\smallskip
\noindent\textbf{\textsc{SybilEdgeTR}.} \textsc{SybilEdgeTR} is a simplified version of \textsc{SybilEdge} that uses only the \emph{target response} component (eq. \ref{responsecomponent}), and not the \emph{target selection} component, so a new user's choice of targets does not affect the posterior probability she is fake (i.e., \textsc{SybilEdgeTR} is \textsc{SybilEdge} with $\sigma$$\rightarrow$$\infty$, see Section \ref{ssec:choosing_phisigma}). \textsc{SybilEdgeTR} probes how much of \textsc{SybilEdge's} performance is due to \emph{target response} alone.

\smallskip 
\noindent\textbf{\textsc{Results}.} Fig. \ref{FBAUC} plots AUC for groups of these new users partitioned by the number of friend requests they sent. \textsc{SybilEdge} and \textsc{SybilEdgeTR} improve in AUC as new users send more friend requests and converge to AUC's of $0.91$ and $0.80$, respectively, for all users who send more than $15$ friend requests. We note that \textsc{SybilEdge}'s high AUC values here mean that it successfully detected even those new users who joined Facebook on the last of the 7 days in the test set (i.e. who were only 1-day old at detection time). This evaluation is (to our knowledge) the first demonstration that a graph-based algorithm can detect fakes given just the small set of friend requests they attempt in their first days of activity.

We also manually inspected \textsc{SybilEdge}'s errors, and we found that similarly to ~\citep{xue2013votetrust}, the class of `false positives' among new users who sent more than $15$ requests reveals many `real-but-spammy' users who abused friend recommendations by sending many unwanted requests. Thus, as in ~\citep{xue2013votetrust}, we conclude that \textsc{SybilEdge}'s `false positives' can actually be desirable outputs.

We also note that, in contrast to some previous evaluations of graph-based algorithms on other social networks, the class of new fake Facebook accounts detected by \textsc{SybilEdge} cannot easily be distinguished by basic network statistics such as reject rates. For example, the authors of \textsc{VoteTrust} note that during their evaluation on the Renren network, fakes were distinguishable by their low average acceptance rate of 0.2 versus 0.8 for real users ~\citep{xue2013votetrust}. In contrast, reject rates yield AUC generally under 0.7 for the class of new users on Facebook. Thus, we conclude that \textsc{SybilEdge} was able to elicit much more information from a new user's sparse friendships by leveraging the differences in targets' selections and responses.


\subsection{SybilEdge vs. state-of-the-art algorithms}
\label{ssec:realdata_estonia}
We also compare \textsc{SybilEdge} to state-of-the-art benchmark algorithms on a Facebook network. Because benchmark algorithms have greater computational complexity than \textsc{SybilEdge} (see Section \ref{ssec:complexity}), we restrict the Facebook network in this evaluation to all users in a single country with roughly $1$ million users. This restriction improves computational feasibility of the benchmarks, and it enables us to use their authors' publicly available code implementations for the sake of experimental transparency (see~\cite{gong2014sybilbelief, wang2017sybilscar, wang2018structure}).

We compare \textsc{SybilEdge} to: 

\smallskip
\noindent \textbf{SybilRank.} \textsc{SybilRank} \cite{yang2012sybilrank} is a state-of-the-art random walk based algorithm. Unlike \textsc{SybilEdge}, \textsc{SybilRank} uses only the graph of accepted friend requests and a set of known real users (nodes). As in \cite{yang2012sybilrank}, we run \textsc{SybilRank} for log2$(|V|)$ iterations.

\smallskip
\noindent \textbf{SybilBelief.} \textsc{SybilBelief}  \cite{gong2014sybilbelief} is a state-of-the-art loopy belief propagation algorithm. \textsc{SybilBelief} uses the friendship graph of accepted friend requests and both known real users and known fakes. As in \cite{gong2014sybilbelief}, we run \textsc{SybilBelief} with edge weights set to $0.9$.

\begin{figure}
\centering
    \includegraphics[width=0.64\linewidth]{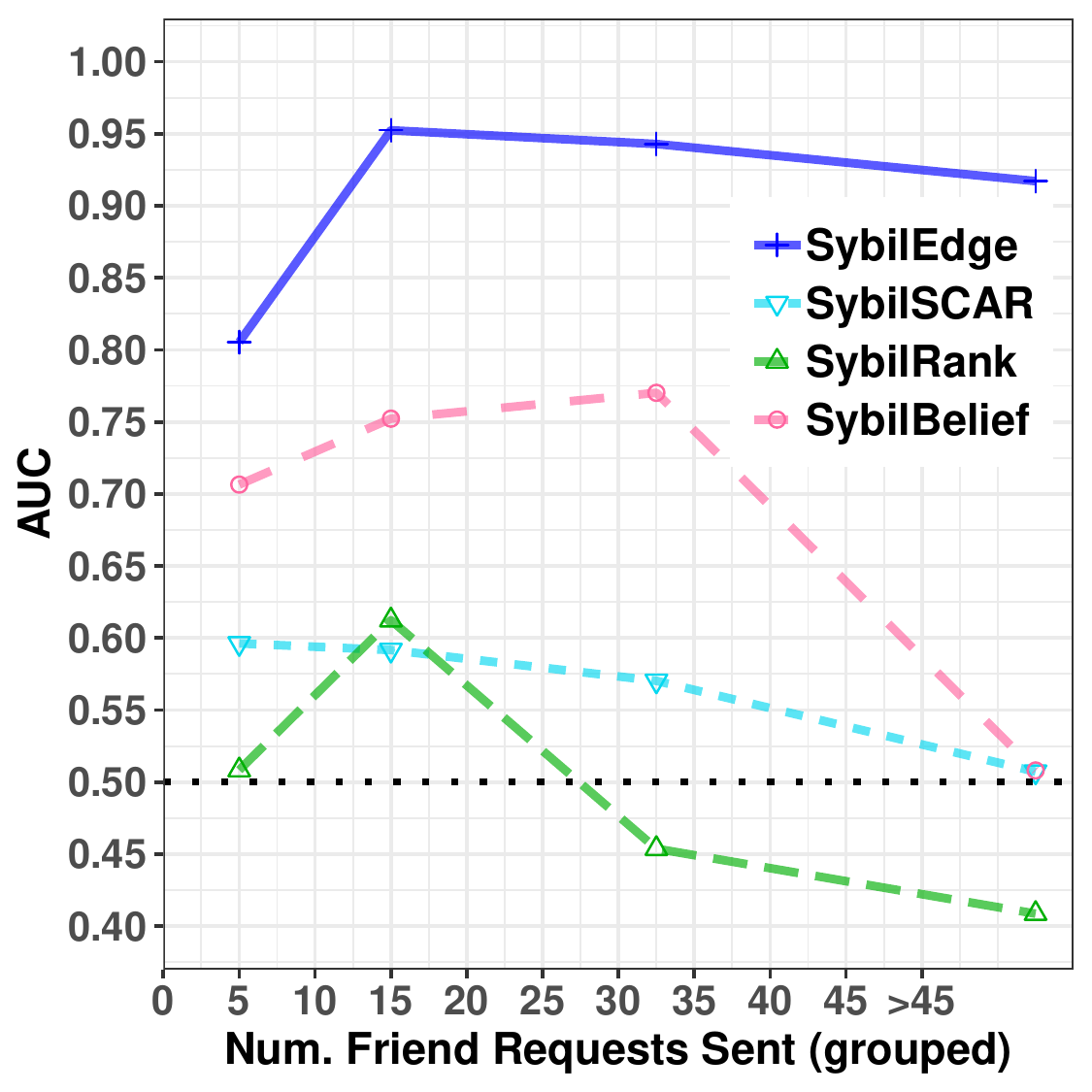}
\caption{Performance of \textsc{SybilEdge} (dark blue) vs. state-of-the-art benchmarks on new Facebook users partitioned by the number of friend requests they sent: $[[0,10], [11,20], [21,45], [46,\infty]]$.}
\Description[\textsc{SybilEdge} vs. benchmarks on the Facebook network]{Plot of AUC vs num. friend requests sent. \textsc{SybilEdge} converges to $AUC>0.9$ after $>10$ requests, whereas benchmarks achieve lower AUC's and inconsistent performance.} 
\label{FBAUC_benchmarks_estonia}
\end{figure}

\smallskip
\noindent \textbf{SybilSCAR.} \textsc{SybilSCAR}  \cite{wang2017sybilscar, wang2018structure} is a recent probabilistic algorithm. \textsc{SybilSCAR} uses the graph of accepted friend requests and both known real users and known fakes. We run both versions of this algorithm: \textsc{SybilSCAR-C} with all weights equal to half the inverse of the average degree as in \cite{wang2018structure}, and user-degree weighted \textsc{SybilSCAR-D}. Each point in Fig.~\ref{FBAUC_benchmarks_estonia} reports the higher of their two AUC's.

\smallskip
\noindent\textbf{Results.}
Fig. \ref{FBAUC_benchmarks_estonia} plots each algorithm's AUC for groups of new users partitioned by the number of friend requests they sent.\footnote{We note that Fig. \ref{FBAUC_benchmarks_estonia} uses fewer partitions than Fig. \ref{FBAUC} to ensure each partition still has sufficient new fake accounts for evaluation on this one-country Facebook network.} Overall, \textsc{SybilEdge} consistently outperforms all benchmarks regardless of the number of friend requests new users sent. Specifically, whereas \textsc{SybilEdge} achieves AUC$>$$0.91$ on all new users who have sent more than 10 friend requests, the best performing benchmark, \textsc{SybilBelief}, achieves a maximum AUC of $0.77$, and its performance degrades to no-better-than-random for new users who send $>$$45$ friend requests. Further investigation suggests that benchmarks' poor performance is largely due to the fact that some new fake users violate the homophily assumption and connect to many indiscriminately accepting real users, and the subset of new fake users who send the most friend requests (for whom the benchmarks' performance is lowest---see rightmost points in Fig. \ref{FBAUC_benchmarks_estonia}) are particularly likely to do so. In these cases, \textsc{SybilRank} tends to rank these fake users in particular as more likely to be real than low-degree real users (resulting in a low or even negative AUC), and \textsc{SybilBelief} and \textsc{SybilSCAR} tend to `over-propagate' known real users' labels via these connections such that the majority of new users converge to identical `$100\%$ real' posteriors, resulting in AUC of $0.5$.

\begin{figure}
\centering
    \includegraphics[width=0.64\linewidth]{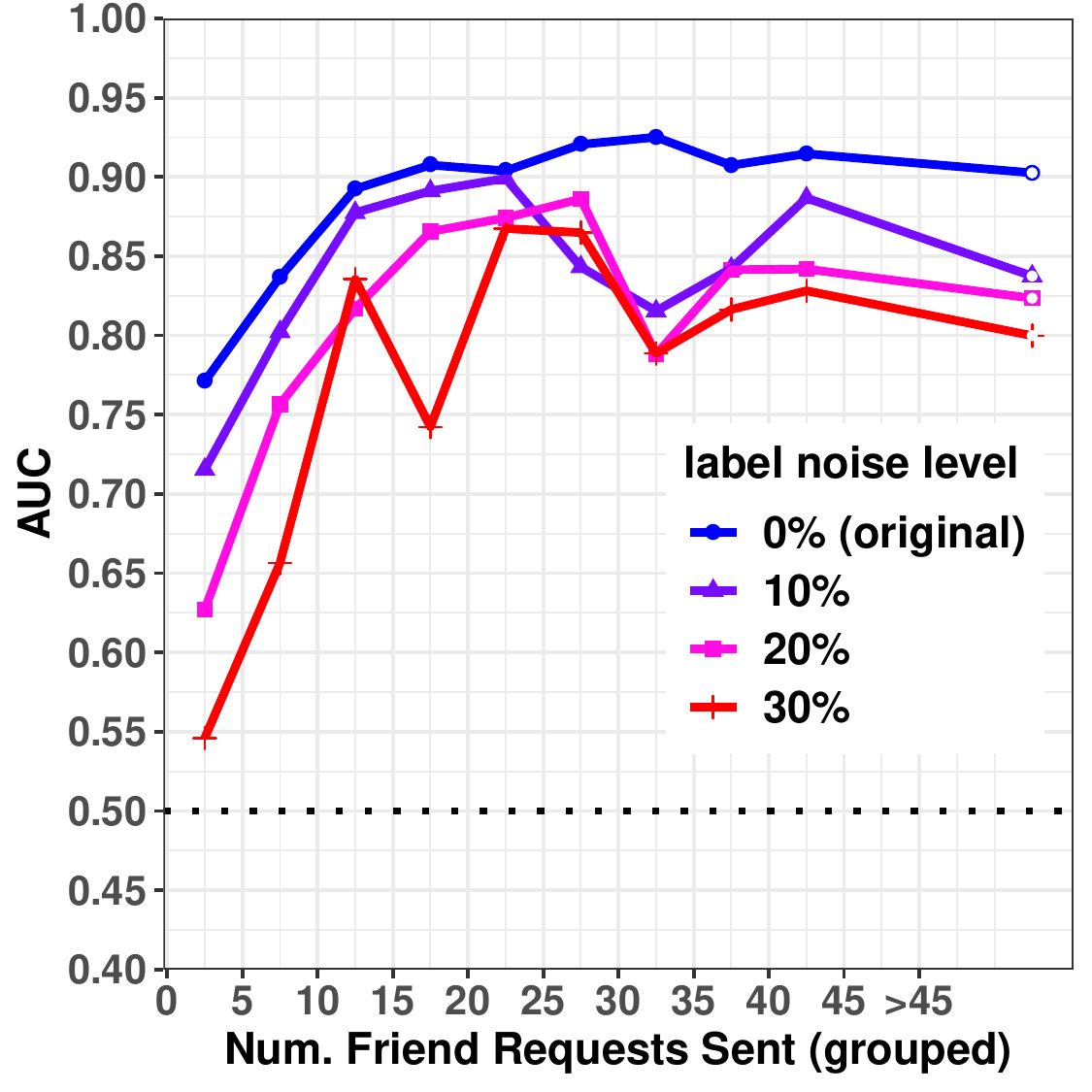}
 \caption{\textsc{SybilEdge} performance under $0\%, 10\%, 20\%,$ and $30\%$ label noise on new global Facebook users partitioned by the number of friend requests they sent: $[[0,5], [6,10],\hdots,[41,45],[46,\infty]]$.}
 \Description[\textsc{SybilEdge} performance under label noise on the Facebook network]{Plot of AUC vs num. friend requests sent under $0\%, 10\%, 20\%,$ and $30\%$ label noise. Under $30\%$ label noise, \textsc{SybilEdge} converges to $AUC>0.8$ after $>21$ requests.} 
\label{FBAUC_NOISY}
\end{figure}

\section{Robustness: label noise}
\label{sec:labelnoise} Robustness to label noise in the training data is a desirable and well-studied property of sybil detection algorithms \cite{gong2014sybilbelief, wang2017sybilscar}. To test the robustness of \textsc{SybilEdge} to noise in a realistic setting, we repeat the evaluation of \textsc{SybilEdge} on the global Facebook network dataset (Fig. \ref{FBAUC}), but randomly flip up to $30$\% of known real and fake users' $\{$fake, real$\}$ labels in the training data we use to compute each target's rates. Fig. \ref{FBAUC_NOISY} plots \textsc{SybilEdge}'s performance on the global Facebook network with various levels of added label noise. Note that even with $30\%$ of added label noise, \textsc{SybilEdge} still converges to $>$$0.80$ AUC on new users who have sent more than $20$ friend requests. We therefore conclude that \textsc{SybilEdge} applies well even to social networks where training labels are known with significantly less confidence than they are at Facebook.

\section{Robustness: behaviors \& prevalence}
\label{sec:Experiments}
Our goal in this section is to show that \textsc{SybilEdge}'s performance advantage is also robust to conditions that differ from the current Facebook network---specifically, to (i) several different ways fakes can select targets for their friend requests, and (ii) greater prevalence of fake accounts in the network. To accomplish this, we designed a variety of synthetic friend request networks to capture a variety of ways fake users can choose targets for their requests. For each synthetic network, we then used real Facebook user data to realistically simulate how Facebook users would respond (accept/reject). Across these simulations, \textsc{SybilEdge} still rapidly converged to detect fakes after they sent only a small number of friend requests regardless of how they selected targets for these requests or their overall prevalence in the network. In all cases, \textsc{SybilEdge} outperformed state-of-the-art graph-based algorithms, whose performance changed markedly depending on how fake users chose targets, and who struggled to detect both low-degree fakes and fakes who succeed in friending less discriminating real users. 

\subsection{Robustness simulations setup}
In each simulation, we set $n=10000$ nodes (users) and randomly select 5\% of them to be fakes, which matches Facebook's global fraction \cite{Standards2019} of fake users (we later increase this to 10\% to probe robustness to a greater prevalence of fake accounts). We randomly select 80\% of nodes to have known fake/real labels and 20\% to have unknown labels. This reflects a realistic `difficult case' of a community where a full 20\% of users are new. We then generate synthetic digraphs of friend requests using a variety of random graph models parameterized by Facebook data. This set of synthetic digraphs is selected to encompass a variety of possible strategies that fakes may deploy ranging from randomly targeting real users to preferentially targeting high-degree users or even users who have previously accepted friend requests from other fakes. For each friend request, we then draw an `accept' or `reject' based on mapping the simulated recipient to an actual Facebook user's accept rates for fakes/reals, which ensures that our simulated users' behaviors are consistent with actual Facebook users. 

\smallskip
\noindent \textbf{Benchmark algorithms} We run \textsc{SybilEdge} and each benchmark algorithm from Section \ref{sec:evaluations} on these graphs to classify the `unknown' $20\%$ of users (test set). We also include an additional benchmark:

\smallskip
\noindent \textbf{VoteTrust.} While we did not run \textsc{VoteTrust} \cite{xue2013votetrust} on the Facebook network (Section \ref{ssec:realdata_estonia}) because it requires significant additional data\footnote{Specifically, as described in Section \ref{ssec:related}, aside from the friendship graph and 3-month sample of users' friend requests we use to train \textsc{SybilEdge} and other benchmarks, \textsc{VoteTrust} also requires complete older historical data on friend requests, which is not among the data that is considered readily accessible for analysis.}, we include it in our simulations. \textsc{VoteTrust} is an interesting benchmark because it is a random walk based algorithm, but like \textsc{SybilEdge}, \textsc{VoteTrust} uses the directed graph of friend requests, accepts, and rejects. \textsc{VoteTrust} detects fakes by propagating trust from known real nodes via random walks, then aggregating accepts/rejects of unknown users' requests weighted by their targets' trust scores.

\begin{figure*}
\centering
\includegraphics[height=0.412\textwidth]{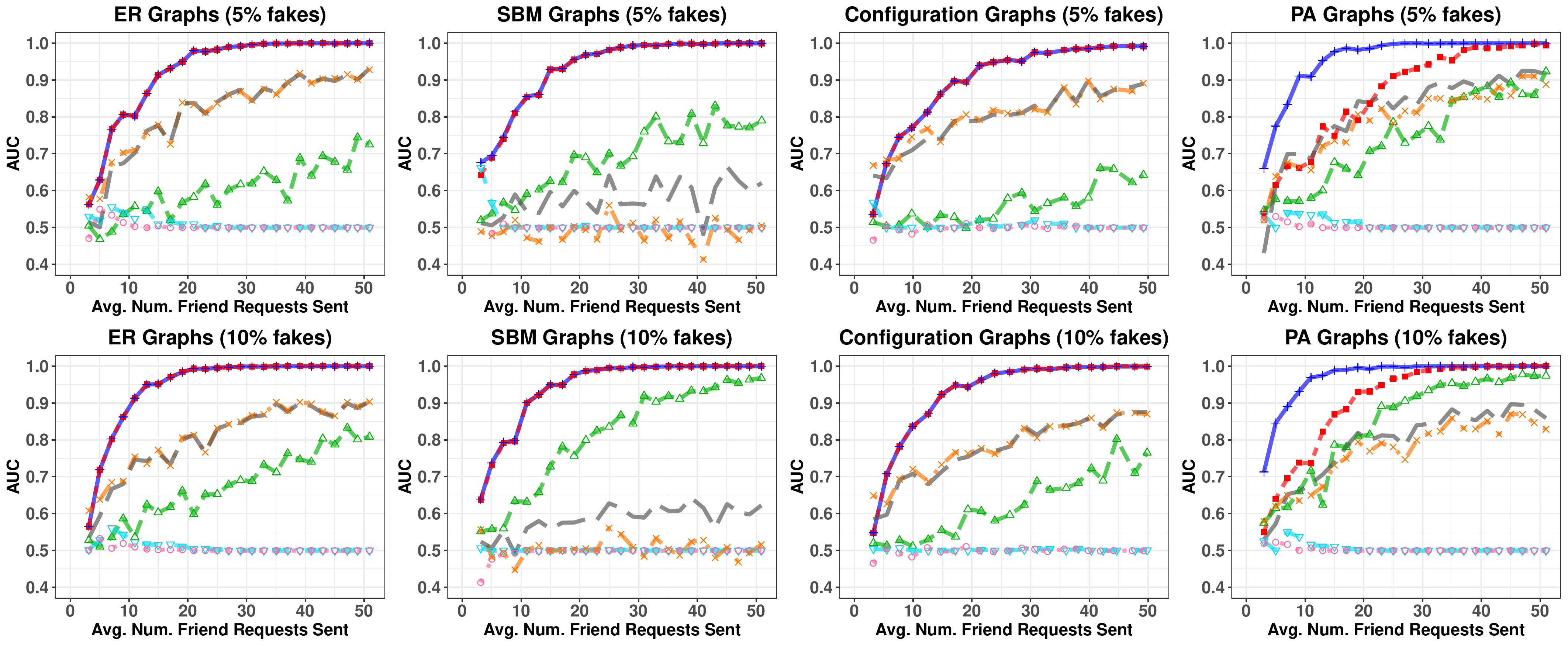}
\includegraphics[width=0.35\textwidth]{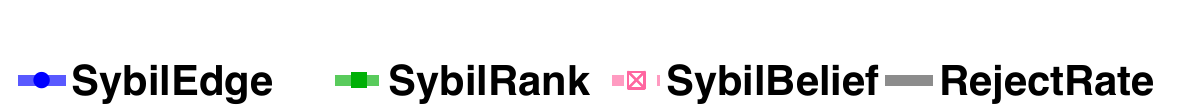}\\
\includegraphics[width=0.24\textwidth]{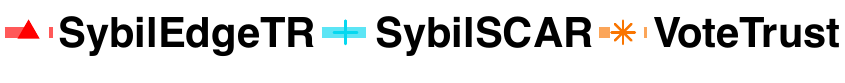}
\caption{Performance of \textsc{SybilEdge} \ (blue) and \textsc{SybilEdgeTR} \ (red) versus benchmarks on random graphs with 5\% fakes (top row) or 10\% fakes (bottom row). Each point represents the AUC of one algorithm on a graph generated with one parameter combination. Different combinations yield graphs with different avg. number of friend requests sent by each user.}
\Description[\textsc{SybilEdge} vs. benchmarks on 8 random graphs]{8 plots of AUC vs avg. num. friend requests sent. \textsc{SybilEdge} converges to $AUC>0.95$ after $\le20$ requests on all plots, whereas benchmarks achieve lower AUC's and inconsistent performance.} 
\label{fig:rand_graphs}
\end{figure*}

\subsection{Generating friend request graphs}
 First, we generate synthetic friend request graphs using various models, each parameterized by Facebook data, which capture various ways fakes can choose their targets. For each graph model, we vary the input parameters to produce a range of graphs with various average out-degrees (number of friend requests sent) from $1$ to $50$.

\smallskip
\noindent \textbf{Erd\H{o}s R\'{e}nyi }(n=10000). We generate friend request graphs using the directed Erd\H{o}s R\'{e}nyi model. We vary the probability $p$ of an edge to yield a range of graphs where nodes' expected number of friend requests varies from 1 to 50 (i.e. $\frac{1}{n-1}\le p \le \frac{50}{n-1}$). These graphs capture a scenario where nodes send friend requests to targets chosen uniformly at random, but targets accept requests as in observed Facebook behavior (see Section \ref{ssec:accepts} below).

\smallskip
\noindent\textbf{FB-parameterized configuration model $(n$=$10000)$.} In practice, some users receive many more friend requests than others. To capture this in a realistic manner, we design directed configuration graphs by mapping each node uniformly at random to an observed Facebook user's count of actual friend requests. We then use each user's count as both her in-degree and out-degree distributions. The resulting graphs capture the scenario where we see a realistic distribution of friend requests, but fakes are careful not to betray their identities by sending many more requests than they receive. 

%
\smallskip
\noindent \textbf{FB-parameterized stochastic block model $(n$=$10000)$.} In practice, real users are much more likely to send requests to other real users than to fakes. We capture this by generating directed SBM graphs of friend requests with two clusters: one of fakes and one of reals. We set the probability of a friend request within- or across-clusters (the edge probability matrix $P_{2,2}$) to the observed ratios at which fakes/reals send requests to fakes/reals on Facebook. 

\smallskip
\noindent \textbf{FB-parameterized preferential attachment $(n$=$10000)$.} In practice, we observe that many fakes preferentially target users who have already been targeted by other fakes (see Fig. \ref{fig:motivation_plots}). To capture this, we design preferential attachment graphs of friend requests. First, we randomly map each simulated fake user uniformly at random to an actual observed fake Facebook user's receive counts from fake/real senders, and we map each simulated real user to corresponding data from a real Facebook user. We these counts as the preferential attachment process weights $\alpha_{fake,j}$ and $\alpha_{real,j}$, i.e., the \emph{a priori} probability that each fake or real user, respectively, will \emph{send} a friend request to target $j$. We then run the classic $k$-out preferential attachment algorithm until all nodes send $k$ friend requests, and we generate a range of graphs with $k=1$ to $50$.

\subsection{Modeling request acceptances/rejections}
\label{ssec:accepts}
After generating a friend request digraph in each simulation, we generate the corresponding `accept' or `reject' for each request as follows: First, we map each simulated target node to a tuple of Facebook data describing a randomly selected Facebook user's historical rates at which she accepted requests from real users and fakes, respectively.\footnote{We note that, due to the fact that millions of users have identical rates, this information is not identifying.} Here, we are careful to map each simulated fake to an actual fake Facebook user's rates and each simulated real user to an actual real Facebook users' rates.\footnote{For the preferential attachment graphs, we are careful to maintain the same mapping as during graph synthesis.} We use these rates as Bernoulli weights to draw `accepts' or `rejects' for her simulated friend requests from real users and fakes, respectively. This process synthesizes realistic `accepts' and `rejects' that match actual Facebook user-level distributions from fake and real accounts.

\subsection{Robustness simulation results}
Fig. \ref{fig:rand_graphs} plots each algorithm's AUC versus the average user's count of friend requests sent (i.e. out-degree) for each graph model. Note that regardless of how fake users selected their targets, \textsc{SybilEdge} consistently achieved near-perfect classification after observing an \emph{average} of 20 friend requests from each user (of which $\sim$4 were sent by unknown users and thus excluded from training). Thus, after training on $\sim$$16$ edges per known user, \textsc{SybilEdge} classified new fakes almost perfectly, \emph{including those who sent only a couple of requests}, across all graph models. This suggests that \textsc{SybilEdge}'s strong performance on the real Facebook network (Section \ref{sec:real_results}) is quite robust to different ways fakes can select targets.

\textsc{SybilEdge} also reaped an additional performance advantage over \textsc{SybilEdgeTR} in preferential attachment graphs, as in these graphs fakes chose targets differently than real users. Per Section \ref{sec:real_results} and Fig. \ref{FBAUC}, this is consistent with \textsc{SybilEdge}'s performance advantage on the real Facebook network.

In contrast, the performance of all benchmark algorithms was markedly inconsistent across the different graph models, and none matched the performance of \textsc{SybilEdge} on any graph model. We inspected their errors and found that, as with evaluations on real data (Section \ref{ssec:realdata_estonia}), benchmarks' poor performance was largely due to the fact that new (simulated) users' sparse connections were insufficient to realize the homophily assumption. 
Also, as in the evaluations on real data, some real users accepted requests indiscriminately from many fakes, causing \textsc{SybilSCAR} and \textsc{SybilBelief} to `over-propagate' known real users' labels out to other fakes, which resulted in many misclassifications. Additionally, all benchmarks struggled to distinguish fakes from low-degree real users.

Finally, \textsc{SybilEdge}'s performance actually \emph{improved} slightly when we increased the fraction of fake accounts in the data from 5\% (\emph{Fig. \ref{fig:rand_graphs} top row}) to 10\% (\emph{bottom row}). This is because the increase in the fraction of fake users improves balance such that a greater fraction of targets in the training data receive requests from known fake users, so \textsc{SybilEdge} can better estimate targets' receive rates and accept rates for fakes when there have been fewer requests overall. This suggests that \textsc{SybilEdge}'s performance is quite robust to even a marked increase in the current fraction of fake accounts.

\section{Conclusion}
\label{sec:conclusion}
We presented \textsc{SybilEdge}, a social-graph-based algorithm for the detection of new fake accounts on social networks. The class of new fakes has traditionally been overlooked by social-graph-based algorithms, which leverage network-structural differences to identify long-tenured fakes. However, we have shown it is possible to detect new fakes by leveraging small individual-level differences in how new fakes interact with other users, and how these users in turn react to new fakes. Because early detection limits the harm that such accounts can inflict, the development of such techniques is a promising new area for impactful research.

\bibliographystyle{ACM-Reference-Format}
\bibliography{fakes_new_bib}


\begin{thebibliography}{32}


\ifx \showCODEN    \undefined \def \showCODEN     #1{\unskip}     \fi
\ifx \showDOI      \undefined \def \showDOI       #1{#1}\fi
\ifx \showISBNx    \undefined \def \showISBNx     #1{\unskip}     \fi
\ifx \showISBNxiii \undefined \def \showISBNxiii  #1{\unskip}     \fi
\ifx \showISSN     \undefined \def \showISSN      #1{\unskip}     \fi
\ifx \showLCCN     \undefined \def \showLCCN      #1{\unskip}     \fi
\ifx \shownote     \undefined \def \shownote      #1{#1}          \fi
\ifx \showarticletitle \undefined \def \showarticletitle #1{#1}   \fi
\ifx \showURL      \undefined \def \showURL       {\relax}        \fi
\providecommand\bibfield[2]{#2}
\providecommand\bibinfo[2]{#2}
\providecommand\natexlab[1]{#1}
\providecommand\showeprint[2][]{arXiv:#2}

\bibitem[\protect\citeauthoryear{Al-Qurishi, Al-Rakhami, Alamri, Alrubaian,
  Rahman, and Hossain}{Al-Qurishi et~al\mbox{.}}{2017}]%
        {al2017sybilreview}
\bibfield{author}{\bibinfo{person}{Muhammad Al-Qurishi},
  \bibinfo{person}{Mabrook Al-Rakhami}, \bibinfo{person}{Atif Alamri},
  \bibinfo{person}{Majed Alrubaian}, \bibinfo{person}{Sk~Md~Mizanur Rahman},
  {and} \bibinfo{person}{M~Shamim Hossain}.} \bibinfo{year}{2017}\natexlab{}.
\newblock \showarticletitle{Sybil defense techniques in online social networks:
  a survey}.
\newblock \bibinfo{journal}{\emph{IEEE Access}}  \bibinfo{volume}{5}
  (\bibinfo{year}{2017}), \bibinfo{pages}{1200--1219}.
\newblock


\bibitem[\protect\citeauthoryear{Alvisi, Clement, Epasto, Lattanzi, and
  Panconesi}{Alvisi et~al\mbox{.}}{2013}]%
        {alvisi2013sok}
\bibfield{author}{\bibinfo{person}{Lorenzo Alvisi}, \bibinfo{person}{Allen
  Clement}, \bibinfo{person}{Alessandro Epasto}, \bibinfo{person}{Silvio
  Lattanzi}, {and} \bibinfo{person}{Alessandro Panconesi}.}
  \bibinfo{year}{2013}\natexlab{}.
\newblock \showarticletitle{Sok: The evolution of sybil defense via social
  networks}. In \bibinfo{booktitle}{\emph{2013 ieee symposium on security and
  privacy}}. IEEE, \bibinfo{pages}{382--396}.
\newblock


\bibitem[\protect\citeauthoryear{Boshmaf, Logothetis, Siganos, Ler{\'\i}a,
  Lorenzo, Ripeanu, and Beznosov}{Boshmaf et~al\mbox{.}}{2015a}]%
        {boshmaf2015integro}
\bibfield{author}{\bibinfo{person}{Yazan Boshmaf}, \bibinfo{person}{Dionysios
  Logothetis}, \bibinfo{person}{Georgos Siganos}, \bibinfo{person}{Jorge
  Ler{\'\i}a}, \bibinfo{person}{Jose Lorenzo}, \bibinfo{person}{Matei Ripeanu},
  {and} \bibinfo{person}{Konstantin Beznosov}.}
  \bibinfo{year}{2015}\natexlab{a}.
\newblock \showarticletitle{Integro: Leveraging Victim Prediction for Robust
  Fake Account Detection in OSNs.}. In \bibinfo{booktitle}{\emph{NDSS}},
  Vol.~\bibinfo{volume}{15}. \bibinfo{pages}{8--11}.
\newblock


\bibitem[\protect\citeauthoryear{Boshmaf, Ripeanu, Beznosov, and
  Santos-Neto}{Boshmaf et~al\mbox{.}}{2015b}]%
        {boshmaf2015thwarting}
\bibfield{author}{\bibinfo{person}{Yazan Boshmaf}, \bibinfo{person}{Matei
  Ripeanu}, \bibinfo{person}{Konstantin Beznosov}, {and}
  \bibinfo{person}{Elizeu Santos-Neto}.} \bibinfo{year}{2015}\natexlab{b}.
\newblock \showarticletitle{Thwarting fake OSN accounts by predicting their
  victims}. In \bibinfo{booktitle}{\emph{Proceedings of the 8th ACM Workshop on
  Artificial Intelligence and Security}}. ACM, \bibinfo{pages}{81--89}.
\newblock


\bibitem[\protect\citeauthoryear{Cao, Sirivianos, Yang, and Munagala}{Cao
  et~al\mbox{.}}{2015}]%
        {cao2015combating}
\bibfield{author}{\bibinfo{person}{Qiang Cao}, \bibinfo{person}{Michael
  Sirivianos}, \bibinfo{person}{Xiaowei Yang}, {and} \bibinfo{person}{Kamesh
  Munagala}.} \bibinfo{year}{2015}\natexlab{}.
\newblock \showarticletitle{Combating friend spam using social rejections}. In
  \bibinfo{booktitle}{\emph{2015 IEEE 35th International Conference on
  Distributed Computing Systems}}. IEEE, \bibinfo{pages}{235--244}.
\newblock


\bibitem[\protect\citeauthoryear{Cao and Yang}{Cao and Yang}{2013}]%
        {cao2013sybilfence}
\bibfield{author}{\bibinfo{person}{Qiang Cao} {and} \bibinfo{person}{Xiaowei
  Yang}.} \bibinfo{year}{2013}\natexlab{}.
\newblock \showarticletitle{SybilFence: Improving social-graph-based sybil
  defenses with user negative feedback}.
\newblock \bibinfo{journal}{\emph{arXiv preprint arXiv:1304.3819}}
  (\bibinfo{year}{2013}).
\newblock


\bibitem[\protect\citeauthoryear{Danezis and Mittal}{Danezis and
  Mittal}{2009}]%
        {danezis2009sybilinfer}
\bibfield{author}{\bibinfo{person}{George Danezis} {and}
  \bibinfo{person}{Prateek Mittal}.} \bibinfo{year}{2009}\natexlab{}.
\newblock \showarticletitle{SybilInfer: Detecting Sybil Nodes using Social
  Networks.}. In \bibinfo{booktitle}{\emph{NDSS}}. San Diego, CA,
  \bibinfo{pages}{1--15}.
\newblock


\bibitem[\protect\citeauthoryear{Facebook}{Facebook}{2019}]%
        {Standards2019}
\bibfield{author}{\bibinfo{person}{Facebook}.} \bibinfo{year}{2019}\natexlab{}.
\newblock \bibinfo{booktitle}{\emph{Community Standards}}.
\newblock
\urldef\tempurl%
\url{https://www.facebook.com/communitystandards}
\showURL{%
\tempurl}


\bibitem[\protect\citeauthoryear{Freeman}{Freeman}{2017}]%
        {freeman2017can}
\bibfield{author}{\bibinfo{person}{David~Mandell Freeman}.}
  \bibinfo{year}{2017}\natexlab{}.
\newblock \showarticletitle{Can you spot the fakes?: On the limitations of user
  feedback in online social networks}. In \bibinfo{booktitle}{\emph{Proceedings
  of the 26th International Conference on World Wide Web}}. International World
  Wide Web Conferences Steering Committee, \bibinfo{pages}{1093--1102}.
\newblock


\bibitem[\protect\citeauthoryear{Fu, Xie, Rui, Gong, Sun, and Chen}{Fu
  et~al\mbox{.}}{2017}]%
        {fu2017robust}
\bibfield{author}{\bibinfo{person}{Hao Fu}, \bibinfo{person}{Xing Xie},
  \bibinfo{person}{Yong Rui}, \bibinfo{person}{Neil~Zhenqiang Gong},
  \bibinfo{person}{Guangzhong Sun}, {and} \bibinfo{person}{Enhong Chen}.}
  \bibinfo{year}{2017}\natexlab{}.
\newblock \showarticletitle{Robust spammer detection in microblogs: Leveraging
  user carefulness}.
\newblock \bibinfo{journal}{\emph{ACM Transactions on Intelligent Systems and
  Technology (TIST)}} \bibinfo{volume}{8}, \bibinfo{number}{6}
  (\bibinfo{year}{2017}), \bibinfo{pages}{83}.
\newblock


\bibitem[\protect\citeauthoryear{Gao, Wang, Gong, Kulkarni, Thomas, and
  Mittal}{Gao et~al\mbox{.}}{2018}]%
        {gao2018sybilfuse}
\bibfield{author}{\bibinfo{person}{Peng Gao}, \bibinfo{person}{Binghui Wang},
  \bibinfo{person}{Neil~Zhenqiang Gong}, \bibinfo{person}{Sanjeev~R Kulkarni},
  \bibinfo{person}{Kurt Thomas}, {and} \bibinfo{person}{Prateek Mittal}.}
  \bibinfo{year}{2018}\natexlab{}.
\newblock \showarticletitle{Sybilfuse: Combining local attributes with global
  structure to perform robust sybil detection}.
\newblock \bibinfo{journal}{\emph{arXiv preprint arXiv:1803.06772}}
  (\bibinfo{year}{2018}).
\newblock


\bibitem[\protect\citeauthoryear{Ghosh, Viswanath, Kooti, Sharma, Korlam,
  Benevenuto, Ganguly, and Gummadi}{Ghosh et~al\mbox{.}}{2012}]%
        {ghosh2012understanding}
\bibfield{author}{\bibinfo{person}{Saptarshi Ghosh}, \bibinfo{person}{Bimal
  Viswanath}, \bibinfo{person}{Farshad Kooti}, \bibinfo{person}{Naveen~Kumar
  Sharma}, \bibinfo{person}{Gautam Korlam}, \bibinfo{person}{Fabricio
  Benevenuto}, \bibinfo{person}{Niloy Ganguly}, {and}
  \bibinfo{person}{Krishna~Phani Gummadi}.} \bibinfo{year}{2012}\natexlab{}.
\newblock \showarticletitle{Understanding and combating link farming in the
  twitter social network}. In \bibinfo{booktitle}{\emph{Proceedings of the 21st
  international conference on World Wide Web}}. ACM, \bibinfo{pages}{61--70}.
\newblock


\bibitem[\protect\citeauthoryear{Gong, Frank, and Mittal}{Gong
  et~al\mbox{.}}{2014}]%
        {gong2014sybilbelief}
\bibfield{author}{\bibinfo{person}{Neil~Zhenqiang Gong}, \bibinfo{person}{Mario
  Frank}, {and} \bibinfo{person}{Prateek Mittal}.}
  \bibinfo{year}{2014}\natexlab{}.
\newblock \showarticletitle{Sybilbelief: A semi-supervised learning approach
  for structure-based sybil detection}.
\newblock \bibinfo{journal}{\emph{IEEE Transactions on Information Forensics
  and Security}} \bibinfo{volume}{9}, \bibinfo{number}{6}
  (\bibinfo{year}{2014}), \bibinfo{pages}{976--987}.
\newblock


\bibitem[\protect\citeauthoryear{Jia, Wang, and Gong}{Jia
  et~al\mbox{.}}{2017}]%
        {jia2017random}
\bibfield{author}{\bibinfo{person}{Jinyuan Jia}, \bibinfo{person}{Binghui
  Wang}, {and} \bibinfo{person}{Neil~Zhenqiang Gong}.}
  \bibinfo{year}{2017}\natexlab{}.
\newblock \showarticletitle{Random walk based fake account detection in online
  social networks}. In \bibinfo{booktitle}{\emph{Dependable Systems and
  Networks (DSN), 2017 47th Annual IEEE/IFIP International Conference on}}.
  IEEE, \bibinfo{pages}{273--284}.
\newblock


\bibitem[\protect\citeauthoryear{Mislove, Marcon, Gummadi, Druschel, and
  Bhattacharjee}{Mislove et~al\mbox{.}}{2007}]%
        {mislove2007measurement}
\bibfield{author}{\bibinfo{person}{Alan Mislove}, \bibinfo{person}{Massimiliano
  Marcon}, \bibinfo{person}{Krishna~P Gummadi}, \bibinfo{person}{Peter
  Druschel}, {and} \bibinfo{person}{Bobby Bhattacharjee}.}
  \bibinfo{year}{2007}\natexlab{}.
\newblock \showarticletitle{Measurement and analysis of online social
  networks}. In \bibinfo{booktitle}{\emph{Proceedings of the 7th ACM SIGCOMM
  conference on Internet measurement}}. ACM, \bibinfo{pages}{29--42}.
\newblock


\bibitem[\protect\citeauthoryear{Mislove, Viswanath, Gummadi, and
  Druschel}{Mislove et~al\mbox{.}}{2010}]%
        {mislove2010you}
\bibfield{author}{\bibinfo{person}{Alan Mislove}, \bibinfo{person}{Bimal
  Viswanath}, \bibinfo{person}{Krishna~P Gummadi}, {and} \bibinfo{person}{Peter
  Druschel}.} \bibinfo{year}{2010}\natexlab{}.
\newblock \showarticletitle{You are who you know: inferring user profiles in
  online social networks}. In \bibinfo{booktitle}{\emph{Proceedings of the
  third ACM international conference on Web search and data mining}}. ACM,
  \bibinfo{pages}{251--260}.
\newblock


\bibitem[\protect\citeauthoryear{Mohaisen, Hopper, and Kim}{Mohaisen
  et~al\mbox{.}}{2011}]%
        {mohaisen2011keep}
\bibfield{author}{\bibinfo{person}{Abedelaziz Mohaisen},
  \bibinfo{person}{Nicholas Hopper}, {and} \bibinfo{person}{Yongdae Kim}.}
  \bibinfo{year}{2011}\natexlab{}.
\newblock \showarticletitle{Keep your friends close: Incorporating trust into
  social network-based sybil defenses}. In \bibinfo{booktitle}{\emph{2011
  Proceedings IEEE INFOCOM}}. IEEE, \bibinfo{pages}{1943--1951}.
\newblock


\bibitem[\protect\citeauthoryear{Qi, AlKulaib, and Broniatowski}{Qi
  et~al\mbox{.}}{2018}]%
        {qi2018detecting}
\bibfield{author}{\bibinfo{person}{SiHua Qi}, \bibinfo{person}{Lulwah
  AlKulaib}, {and} \bibinfo{person}{David~A Broniatowski}.}
  \bibinfo{year}{2018}\natexlab{}.
\newblock \showarticletitle{Detecting and characterizing bot-like behavior on
  Twitter}. In \bibinfo{booktitle}{\emph{International Conference on Social
  Computing, Behavioral-Cultural Modeling and Prediction and Behavior
  Representation in Modeling and Simulation}}. Springer,
  \bibinfo{pages}{228--232}.
\newblock


\bibitem[\protect\citeauthoryear{Ramalingam and Chinnaiah}{Ramalingam and
  Chinnaiah}{2018}]%
        {ramalingam2018fakesybilreview}
\bibfield{author}{\bibinfo{person}{Devakunchari Ramalingam} {and}
  \bibinfo{person}{Valliyammai Chinnaiah}.} \bibinfo{year}{2018}\natexlab{}.
\newblock \showarticletitle{Fake profile detection techniques in large-scale
  online social networks: A comprehensive review}.
\newblock \bibinfo{journal}{\emph{Computers \& Electrical Engineering}}
  \bibinfo{volume}{65} (\bibinfo{year}{2018}), \bibinfo{pages}{165--177}.
\newblock


\bibitem[\protect\citeauthoryear{Raykar, Yu, Zhao, Valadez, Florin, Bogoni, and
  Moy}{Raykar et~al\mbox{.}}{2010}]%
        {raykar2010learning}
\bibfield{author}{\bibinfo{person}{Vikas~C Raykar}, \bibinfo{person}{Shipeng
  Yu}, \bibinfo{person}{Linda~H Zhao}, \bibinfo{person}{Gerardo~Hermosillo
  Valadez}, \bibinfo{person}{Charles Florin}, \bibinfo{person}{Luca Bogoni},
  {and} \bibinfo{person}{Linda Moy}.} \bibinfo{year}{2010}\natexlab{}.
\newblock \showarticletitle{Learning from crowds}.
\newblock \bibinfo{journal}{\emph{Journal of Machine Learning Research}}
  \bibinfo{volume}{11}, \bibinfo{number}{Apr} (\bibinfo{year}{2010}),
  \bibinfo{pages}{1297--1322}.
\newblock


\bibitem[\protect\citeauthoryear{Subrahmanian, Azaria, Durst, Kagan, Galstyan,
  Lerman, Zhu, Ferrara, Flammini, and Menczer}{Subrahmanian
  et~al\mbox{.}}{2016}]%
        {subrahmanian2016darpa}
\bibfield{author}{\bibinfo{person}{VS Subrahmanian}, \bibinfo{person}{Amos
  Azaria}, \bibinfo{person}{Skylar Durst}, \bibinfo{person}{Vadim Kagan},
  \bibinfo{person}{Aram Galstyan}, \bibinfo{person}{Kristina Lerman},
  \bibinfo{person}{Linhong Zhu}, \bibinfo{person}{Emilio Ferrara},
  \bibinfo{person}{Alessandro Flammini}, {and} \bibinfo{person}{Filippo
  Menczer}.} \bibinfo{year}{2016}\natexlab{}.
\newblock \showarticletitle{The DARPA Twitter bot challenge}.
\newblock \bibinfo{journal}{\emph{Computer}} \bibinfo{volume}{49},
  \bibinfo{number}{6} (\bibinfo{year}{2016}), \bibinfo{pages}{38--46}.
\newblock


\bibitem[\protect\citeauthoryear{Wang, Gong, and Fu}{Wang
  et~al\mbox{.}}{2017a}]%
        {wang2017gang}
\bibfield{author}{\bibinfo{person}{Binghui Wang},
  \bibinfo{person}{Neil~Zhenqiang Gong}, {and} \bibinfo{person}{Hao Fu}.}
  \bibinfo{year}{2017}\natexlab{a}.
\newblock \showarticletitle{GANG: Detecting fraudulent users in online social
  networks via guilt-by-association on directed graphs}. In
  \bibinfo{booktitle}{\emph{Data Mining (ICDM), 2017 IEEE International
  Conference on}}. IEEE, \bibinfo{pages}{465--474}.
\newblock


\bibitem[\protect\citeauthoryear{Wang, Jia, Zhang, and Gong}{Wang
  et~al\mbox{.}}{2018a}]%
        {wang2018structure}
\bibfield{author}{\bibinfo{person}{Binghui Wang}, \bibinfo{person}{Jinyuan
  Jia}, \bibinfo{person}{Le Zhang}, {and} \bibinfo{person}{Neil~Zhenqiang
  Gong}.} \bibinfo{year}{2018}\natexlab{a}.
\newblock \showarticletitle{Structure-based sybil detection in social networks
  via local rule-based propagation}.
\newblock \bibinfo{journal}{\emph{IEEE Transactions on Network Science and
  Engineering}} (\bibinfo{year}{2018}).
\newblock


\bibitem[\protect\citeauthoryear{Wang, Zhang, and Gong}{Wang
  et~al\mbox{.}}{2017b}]%
        {wang2017sybilscar}
\bibfield{author}{\bibinfo{person}{Binghui Wang}, \bibinfo{person}{Le Zhang},
  {and} \bibinfo{person}{Neil~Zhenqiang Gong}.}
  \bibinfo{year}{2017}\natexlab{b}.
\newblock \showarticletitle{SybilSCAR: Sybil detection in online social
  networks via local rule based propagation}. In
  \bibinfo{booktitle}{\emph{INFOCOM 2017-IEEE Conference on Computer
  Communications, IEEE}}. IEEE, \bibinfo{pages}{1--9}.
\newblock


\bibitem[\protect\citeauthoryear{Wang, Zhang, and Gong}{Wang
  et~al\mbox{.}}{2018b}]%
        {wang2018sybilblind}
\bibfield{author}{\bibinfo{person}{Binghui Wang}, \bibinfo{person}{Le Zhang},
  {and} \bibinfo{person}{Neil~Zhenqiang Gong}.}
  \bibinfo{year}{2018}\natexlab{b}.
\newblock \showarticletitle{Sybilblind: Detecting fake users in online social
  networks without manual labels}. In \bibinfo{booktitle}{\emph{International
  Symposium on Research in Attacks, Intrusions, and Defenses}}. Springer,
  \bibinfo{pages}{228--249}.
\newblock


\bibitem[\protect\citeauthoryear{Xue, Yang, Yang, Wang, Chen, and Dai}{Xue
  et~al\mbox{.}}{2013}]%
        {xue2013votetrust}
\bibfield{author}{\bibinfo{person}{Jilong Xue}, \bibinfo{person}{Zhi Yang},
  \bibinfo{person}{Xiaoyong Yang}, \bibinfo{person}{Xiao Wang},
  \bibinfo{person}{Lijiang Chen}, {and} \bibinfo{person}{Yafei Dai}.}
  \bibinfo{year}{2013}\natexlab{}.
\newblock \showarticletitle{Votetrust: Leveraging friend invitation graph to
  defend against social network sybils}. In \bibinfo{booktitle}{\emph{2013
  Proceedings IEEE INFOCOM}}. IEEE, \bibinfo{pages}{2400--2408}.
\newblock


\bibitem[\protect\citeauthoryear{Yang, Harkreader, Zhang, Shin, and Gu}{Yang
  et~al\mbox{.}}{2012b}]%
        {yang2012analyzing}
\bibfield{author}{\bibinfo{person}{Chao Yang}, \bibinfo{person}{Robert
  Harkreader}, \bibinfo{person}{Jialong Zhang}, \bibinfo{person}{Seungwon
  Shin}, {and} \bibinfo{person}{Guofei Gu}.} \bibinfo{year}{2012}\natexlab{b}.
\newblock \showarticletitle{Analyzing spammers' social networks for fun and
  profit: a case study of cyber criminal ecosystem on twitter}. In
  \bibinfo{booktitle}{\emph{Proceedings of the 21st international conference on
  World Wide Web}}. ACM, \bibinfo{pages}{71--80}.
\newblock


\bibitem[\protect\citeauthoryear{Yang, Cao, and Sirivianos}{Yang
  et~al\mbox{.}}{2012a}]%
        {yang2012sybilrank}
\bibfield{author}{\bibinfo{person}{X Yang}, \bibinfo{person}{Q Cao}, {and}
  \bibinfo{person}{M Sirivianos}.} \bibinfo{year}{2012}\natexlab{a}.
\newblock \bibinfo{title}{SybilRank: Aiding the detection of fake accounts in
  large scale social online services}.
\newblock
\newblock


\bibitem[\protect\citeauthoryear{Yang, Wilson, Wang, Gao, Zhao, and Dai}{Yang
  et~al\mbox{.}}{2014}]%
        {yang2014uncovering}
\bibfield{author}{\bibinfo{person}{Zhi Yang}, \bibinfo{person}{Christo Wilson},
  \bibinfo{person}{Xiao Wang}, \bibinfo{person}{Tingting Gao},
  \bibinfo{person}{Ben~Y Zhao}, {and} \bibinfo{person}{Yafei Dai}.}
  \bibinfo{year}{2014}\natexlab{}.
\newblock \showarticletitle{Uncovering social network sybils in the wild}.
\newblock \bibinfo{journal}{\emph{ACM Transactions on Knowledge Discovery from
  Data (TKDD)}} \bibinfo{volume}{8}, \bibinfo{number}{1}
  (\bibinfo{year}{2014}), \bibinfo{pages}{2}.
\newblock


\bibitem[\protect\citeauthoryear{Yu, Gibbons, Kaminsky, and Xiao}{Yu
  et~al\mbox{.}}{2008}]%
        {yu2008sybillimit}
\bibfield{author}{\bibinfo{person}{Haifeng Yu}, \bibinfo{person}{Phillip~B
  Gibbons}, \bibinfo{person}{Michael Kaminsky}, {and} \bibinfo{person}{Feng
  Xiao}.} \bibinfo{year}{2008}\natexlab{}.
\newblock \showarticletitle{Sybillimit: A near-optimal social network defense
  against sybil attacks}. In \bibinfo{booktitle}{\emph{2008 IEEE Symposium on
  Security and Privacy (sp 2008)}}. IEEE, \bibinfo{pages}{3--17}.
\newblock


\bibitem[\protect\citeauthoryear{Yu, Kaminsky, Gibbons, and Flaxman}{Yu
  et~al\mbox{.}}{2006}]%
        {yu2006sybilguard}
\bibfield{author}{\bibinfo{person}{Haifeng Yu}, \bibinfo{person}{Michael
  Kaminsky}, \bibinfo{person}{Phillip~B Gibbons}, {and}
  \bibinfo{person}{Abraham Flaxman}.} \bibinfo{year}{2006}\natexlab{}.
\newblock \showarticletitle{Sybilguard: defending against sybil attacks via
  social networks}. In \bibinfo{booktitle}{\emph{ACM SIGCOMM Computer
  Communication Review}}, Vol.~\bibinfo{volume}{36}. ACM,
  \bibinfo{pages}{267--278}.
\newblock


\bibitem[\protect\citeauthoryear{Zhang, Zhang, Sun, Zhang, and Zhang}{Zhang
  et~al\mbox{.}}{2015}]%
        {zhang2015truetop}
\bibfield{author}{\bibinfo{person}{Jinxue Zhang}, \bibinfo{person}{Rui Zhang},
  \bibinfo{person}{Jingchao Sun}, \bibinfo{person}{Yanchao Zhang}, {and}
  \bibinfo{person}{Chi Zhang}.} \bibinfo{year}{2015}\natexlab{}.
\newblock \showarticletitle{Truetop: A sybil-resilient system for user
  influence measurement on twitter}.
\newblock \bibinfo{journal}{\emph{IEEE/ACM Transactions on Networking}}
  \bibinfo{volume}{24}, \bibinfo{number}{5} (\bibinfo{year}{2015}),
  \bibinfo{pages}{2834--2846}.
\newblock


\end{thebibliography}
\end{document}